\documentclass[fleqn,10pt,twocolumn]{wlscirep}
\usepackage[utf8]{inputenc}
\usepackage[T1]{fontenc}
\usepackage{bm}
\usepackage{comment}
\usepackage{mhchem}
\usepackage{soul} 
\usepackage{amsmath}
 \title{Temporal network restructuring improves control of indecisive collectives}

\author[1]{Tuhin Chakrabortty}
\author[1]{Saad Bhamla}

\affil[1]{Georgia Institute of Technology, USA}

\affil[*]{saadb@chbe.gatech.edu}

\begin{abstract}

Controlling stochastic temporal networks remains an open challenge in control theory. While predictable temporal networks with known future dynamics enhance controllability, real-world networks often exhibit stochasticity and unpredictability, making control harder. Here, we investigate control mechanisms for stochastic temporal networks by analyzing how biological controllers, such as shepherd dogs, manage panicked flocks of sheep. We studied a century-old shepherding competition, the sheepdog trials, where small groups of sheep unpredictably switch between fleeing and following behaviors--effectively forming stochastic temporal networks. Unlike large, cohesive flocks, these small, indecisive flocks are difficult to control, yet skilled dog-handler teams excel at both herding and splitting them (shedding) on demand. Using a stochastic choice model to describe the sheep’s behavioral shifts, we found that trained dogs exploit stochastic indecisiveness, typically seen as an obstacle, as a control tool, enabling both herding and splitting of noisy groups of sheep. Building on these insights, we developed the Indecisive Swarm Algorithm (ISA) for artificial agents and benchmarked its performance against standard approaches, including the Averaging-Based Swarm Algorithm (ASA) and the Leader-Follower Swarm Algorithm (LFSA). ISA minimizes control energy in trajectory-following tasks and outperforms alternatives under noisy conditions. By framing these results within a stochastic temporal network perspective, we demonstrate that even probabilistic knowledge of future dynamics can enhance control efficiency in specific scenarios. These findings establish a framework for managing stochastic temporal networks with applications in noisy, behavior-switching animal collectives, swarm robotics, and opinion dynamics.

\end{abstract}
\begin{document}

\flushbottom
\maketitle

\thispagestyle{empty}









\section*{Introduction}

Emergent collective dynamics, where simple local interactions give rise to complex global behaviors, govern a wide range of systems. Examples include swarm robotics \cite{schranz2020swarm}, animal collectives \cite{sumpter2010collective},  social networks such as opinion dynamics \cite{anderson2019recent}, pedestrians’ movements \cite{helbing2001self}, and vehicular traffic \cite{kessels2019traffic}. Controlling these systems is challenging, as their behaviors often defy traditional control methods \cite{doyle2013feedback,lewis2013cooperative, d2023controlling}. Unlike systems with predictable, linear dynamics, emergent systems are best described as complex networks that require multiscale strategies to address both the microscopic interactions between individual agents  and the macroscopic patterns that emerge at the group level  \cite{d2023controlling}. 

Most of these networked systems introduce additional complexity when individual agents (nodes) switch between different behaviors, leading to temporal restructuring in the network.  Biological collectives and social interactions in humans serve as prime examples of such behavior switching ~\cite{feinerman2018physics, Buhl2006-dw, Jhawar2020-uw,gomez2022intermittent,zha2020opinion} (see SI Section 1 and Table S1 for a full list of behavior-switching systems from ants and locusts to seals and humans). Carrier ants transporting cargo alternate roles between lifters and pullers based on their orientation and the nest's position \cite{feinerman2018physics}, sheep in small flocks randomly switch between leading and following roles \cite{gomez2022intermittent}, and during an epidemic outbreak, humans frequently switch between different interaction partners, facilitating spread of diseases \cite{Koher2019-qo}. These systems highlight the need for control strategies that account for the stochastic and context-dependent nature of individual behavior transitions and their cascading effects on the evolving temporal networks, where edges dynamically reorganize over time \cite{masuda2016guide,humphries2021systematic, Holme2012-tc}.

Recently, it has been shown that temporal restructuring can improve the controllability of a network \cite{li2017fundamental}. Specifically, temporal networks require less time and less energetic cost to be controlled than their static counterparts \cite{Zhang2021-pa, Lebon2024-kw}. This counterintuitive observation relies on the fact that the future dynamics of the network are predictable and are exploited in designing the controls in the previous steps. However, when the switching dynamics are stochastic and unpredictable, temporality can make the control process more energetically demanding compared to a static network \cite{de2021inherent}. Therefore, despite advances in control theory and swarm robotics \cite{lewis2013cooperative, d2023controlling}, managing the dynamics of stochastic temporal networks remains an open challenge, particularly in systems where individual agents exhibit behavior switching.

Predator-prey systems provide a natural framework for studying the challenges of controlling such noisy networks with behavior-switching dynamics. For instance, flocks of starlings confuse raptors by transitioning between complex dynamic patterns. Similarly, large herds of wildebeests intermittently shift between selfish herding and solitary flight when confronted by predators like cheetahs. In response, predators, instead of complex control mechanisms, adopt simplified strategies like focusing on a fixed point in space rather than tracking individual prey \cite{storms2019complex, brighton2022raptors}. This allows them to split vulnerable individuals before leveraging speed and agility to secure their target \cite{ruxton2002living, noauthor_undated-rn, noauthor_youtube}. These examples suggest that effective control of stochastic temporal networks with behavior-switching individuals does not always require precise prediction of behavioral transitions. 

In this work, we analyze such control mechanisms by studying shepherd dogs managing small flocks of sheep in a competition called the sheep-dog trials. Two key features of these competitions make them model systems for investigating control mechanisms in stochastic temporal networks. First,  during these trials, when threatened, panicked sheep oscillate unpredictably and indecisively between fleeing from the dog and following other sheep, forming a stochastic temporal network. Trained shepherd dogs are highly effective at managing these noisy flocks under fluctuating conditions (SI Video 1). Second,  unlike interactions between predators and large herds of animals in the wild, the sheepdog trials competitions provide a controlled environment where the behavior-switching dynamics of the sheep can be observed, quantified, and analyzed (see SI Section 2\&3 for history and competition rules). 

By bridging empirical observations with quantitative modeling to analyze various tasks in the sheepdog trials competition, we find that shepherd dogs utilize the behavior switching in sheep for herding and splitting (shedding) the flocks. Behavior switching dynamics have been previously studied in the context of animal collectives and human societies (voter models) using individual-based stochastic choice models \cite{sardanyes2018noise, Jhawar2020-uw, Biancalani2014-dr, dyson2015onset, redner2019reality}.  In this work, we build on the existing framework to frame sheep-dog trials as a control problem for  "indecisive collectives" — systems where agents stochastically alternate between different behaviors and interaction partners in the presence of an external agent. 

This paper is structured as follows: We begin by exploring the nuances and rules of sheepdog trials. Next, we present a stochastic framework to develop quantitative metrics such as "pressure" and "lightness" that capture the nuanced behavior of sheep. The framework is based on qualitative insights from experienced handlers, and empirical data on sheep-dog dynamics.  We then present a stochastic choice model and the master equation to model the indecisive transitions in sheep movement, comparing our model's predictions with observed dynamics. Building on this, we investigate whether sheep indecisiveness could benefit the dog. Our findings reveal that stochastic indecisiveness can aid the dog in both herding and shedding tasks. Finally, we extend our analysis to develop the Indecisive Swarm Algorithm (ISA), a swarm control strategy inspired by shepherding dynamics. By modeling ISA as a non-reciprocal stochastic temporal network and comparing it against the standard Averaging-Based Algorithm (ASA) and Leader-Follower Swarm Algorithm (LFSA), we demonstrate that for specific control tasks like herding, ISA minimizes control energy requirements.

\section*{Results}
\subsection*{Terminology in Sheep-dog Trials}
\begin{figure*}[h!t]
\centering
    \includegraphics[width = .99\textwidth]{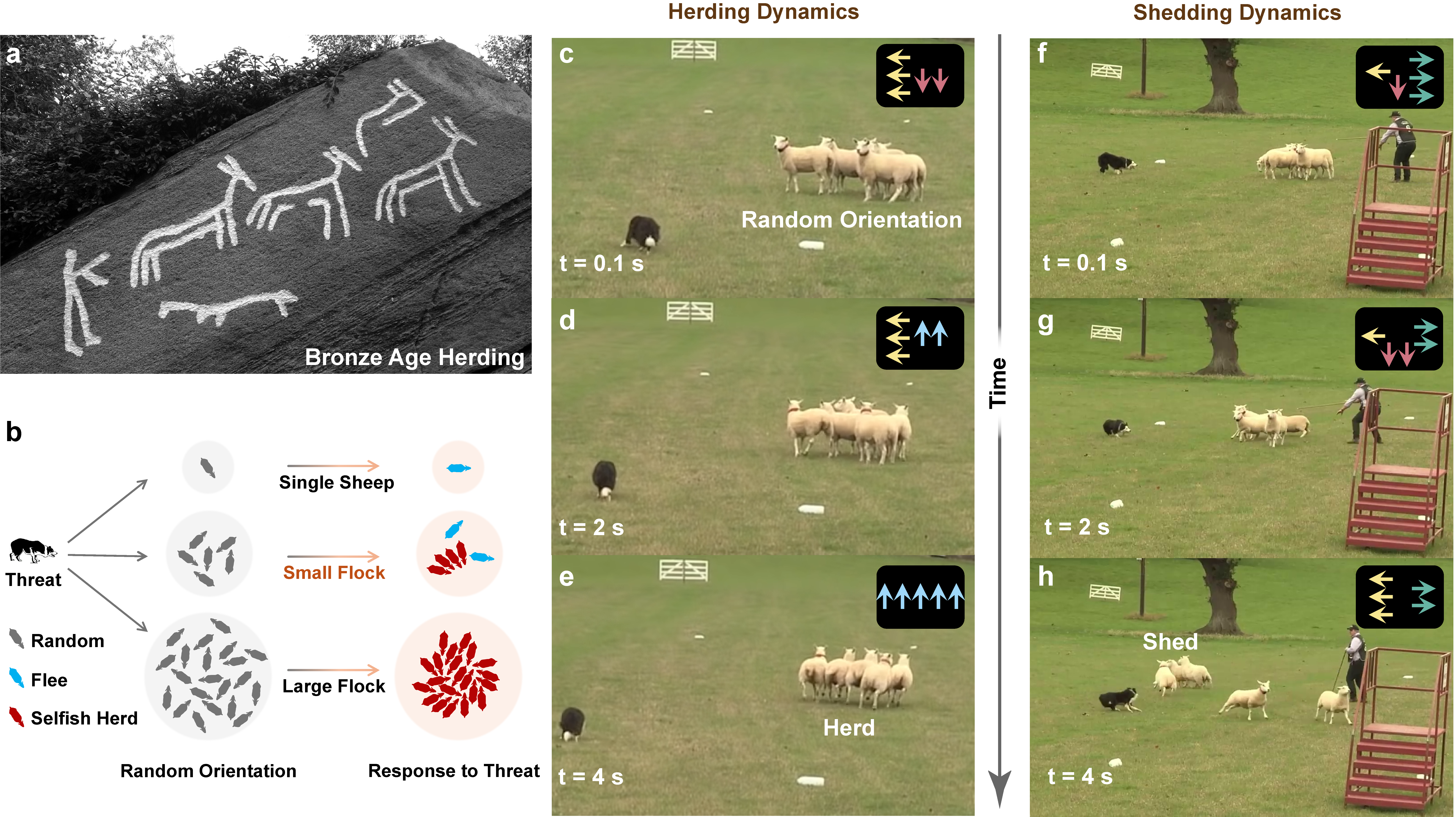}
    \caption{\textbf{Human-Dog-Sheep Interaction in Small Groups} \textbf{a} A Bronze Age rock art panel at Valhaug on Jæren in southwestern Norway showing a shepherd herding a small group of sheep with the help of a dog (Photo Credit: Paul G. Keil)~\cite{keil2015human}. \textbf{b} Transition from single sheep response to large group response: While a single sheep flees under threat, sheep in a large group show selfish herd behavior. Sheep in small groups are highly indecisive and show a stochastic transition between the two behaviors,  making the groups unpredictable.  \textbf{c-e} Dynamics of herding in real sheep-dog system(SI video 3). \textbf{f-h} Dynamics of shedding in real sheep-dog system (SI video 3).}
    \label{fig:Fig1}
    \end{figure*}

Historically, shepherds have  exploited  predator-prey dynamics to control collectives,  leveraging  herding dogs to manage farm animals  as early as 1700 BC (see Figure ~\ref{fig:Fig1}a, SI Section 2)~\cite{grandin2019livestock}. When a solitary sheep encounters  a threat, it flees; in large groups, sheep exhibit selfish herd behavior \cite{King2012-km}. However, in small groups, sheep struggle to choose a survival strategy, indecisively switching between solitary and collective behaviors, creating unpredictability  (Figure ~\ref{fig:Fig1}b and SI Video 1). This unpredictability  led to  the creation of sheep-dog trials, a 100-year-old sporting competition testing a dog's ability to control small sheep groups ($N_s\le5$)~\cite{noauthor_undated-xk}. In these trials, handlers and dogs not only move sheep cohesively (called herding), but also split the groups into subgroups (called shedding), showcasing the dog's skill in managing indecisive collectives (Figure ~\ref{fig:Fig1}c-h, see SI Section 3) ~\cite{keil2015human}.


In sheep-dog trials, qualitative terms such as "pressure" and "lightness" convey the following aspects: pressure refers to the threat perceived by sheep from a dog's actions, such as approaching, barking, or staring, while lightness describes the sheep's responsiveness to these cues \cite{holland2007herding}. Trained dogs apply pressure to herd or shed (split) sheep, and handlers also contribute by exerting pressure through their body posture during shedding. Lightness is a measure of the responsiveness of the sheep. Light sheep respond to minimal pressure but may panic under high pressure, whereas heavy sheep resist until high pressure is applied directly from the front. Assessing sheep lightness early in trials is essential for achieving effective control \cite{holland2007herding}.

\subsection*{Empirical Analysis of Orientation Dynamics to Inform Modeling}
To examine how the control strategies of the dog differ between light and heavy sheep and to translate this nuanced qualitative knowledge into a quantitative framework, we recognize that herding and shedding both involve two sequential steps \cite{early2020sequential}. The first step, which we term the \textit{orientation} step, involves nudging stationary sheep gently to induce directional change without causing panic. The second step, termed the \textit{movement} step, involves increasing pressure to prompt movement (see SI section 4b and SI Video 2 for more details) \cite{keil2015human}.  Our study focuses only on the initial orientation step, isolating it from spatial dynamics such as movement and steric interactions, and considers only the  orientation of sheep relative to the dog. 

In sheep-dog trials, for a group of 5 sheep, a herding state is achieved when all the sheep are oriented away from the dog (Figure \ref{fig:Fig1}e). A shedding state occurs when the sheep divide into two groups of 3 and 2 individuals, with each group oriented perpendicularly away from both the dog and the handler (Figure \ref{fig:Fig1}h). To simplify our analysis, we classify  sheep orientations into 4 directions relative to the dog: directly facing, perpendicular left, perpendicular right, and facing away (See SI Video 3, SI section 4a).  

By quantifying the transitions between these 4 orientations in 21 videos of  sheep-dog trials (See SI Section 4a for details), we observe clear differences in the behavior of light and heavy sheep during herding and shedding. In herding, light sheep quickly achieve a herding state and remain there, with occasional individual escapes. Heavy sheep, in contrast, exhibit intermittent herding states, often aligning orthogonally as a flock to the dog (Figure~\ref{fig:Fig3} a,c Bottom, SI Videos 6 and 7; Table S2). During shedding, light sheep frequently reorient individually, often due to panicking (Figure~\ref{fig:Fig3}b Bottom). Heavy sheep, however, tend to align orthogonally to the dog and handler, synchronously switching between the two perpendicular directions away from the handler and the dog  as a group (Figure \ref{fig:Fig3}b,d Bottom; SI Videos 8 and 9; Table S2). 

The observed differences in behavior between light and heavy sheep, particularly their orientation transitions under varying pressure and lightness conditions, form the basis for defining the parameters in our stochastic model. These insights motivate the development of a quantitative framework to predict and generalize the indecisive behavior-switching dynamics observed in sheep-dog trials.

\begin{figure*}[h!t]
    \centering
    \includegraphics[width = \textwidth]{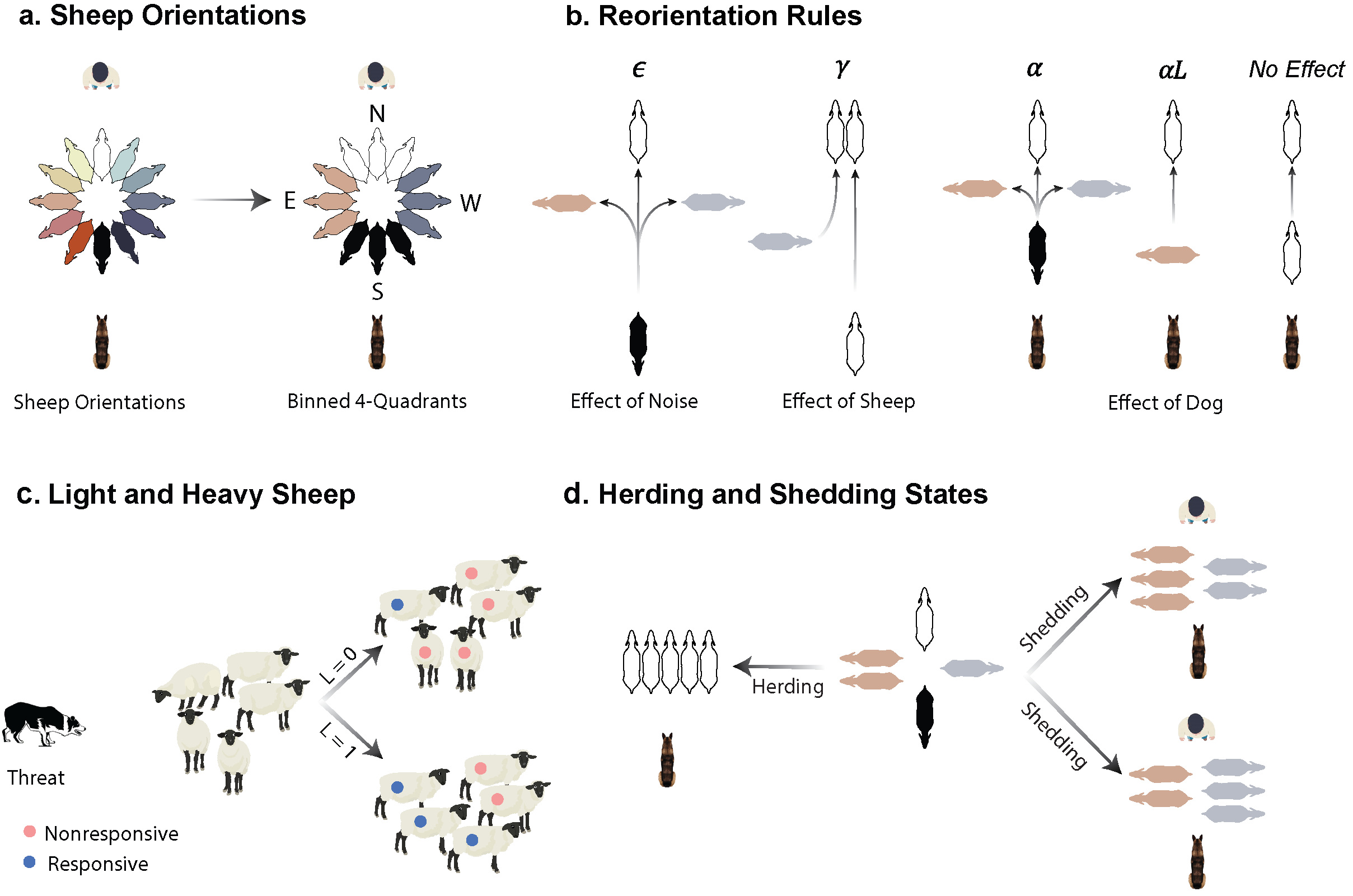}
    \caption{\textbf{Quantitative Framework for Modeling the Orientation Step}  We simplify the model by making two assumptions: \textbf{a.} We only consider the orientation of the sheep and bin the 2D space in 4 directions, allowing us to model sheep as stationary pointers that can reorient in 4 possible directions.  \textbf{b.} Transition rules describe how a sheep changes its direction when influenced by a dog/handler, other sheep, or spontaneously due to random noise with rates $\alpha_{ij}, \gamma$, and $\epsilon$, respectively. The parameter $\alpha_{ij}$ represents the threat from the dog present in direction $j$ on the sheep oriented in direction $i$. When influenced by a dog/handler, sheep's behavior changes depending on their orientation. Sheep facing the dog panic and randomly reorient, sheep perpendicular to the dog reorient to the opposite direction of the dog and sheep oriented away from the dog don't change their orientation.  \textbf{c.} Definition of lightness and responsiveness of sheep. Ideal light sheep with lightness $(L) = 1$ respond to the dog irrespective of their orientation. Ideal heavy sheep with $L = 0$ only respond if they are facing the dog. Sheep with intermediate lightness $0 < L < 1$ have higher responsiveness when facing the dog compared to being perpendicular to the dog \textbf{d.} Description of herding and shedding processes in our model. In herding, the goal is to align all the sheep away from the dog, whereas in shedding, the goal is to divide the group into two subgroups as required (typically into 3 and 2). Shedding involves both the handler and the dog.}   
    \label{fig:Fig2}
    \end{figure*}

\begin{figure*}[htbp]
    \centering
    
 \includegraphics[width = 0.9\textwidth]
 {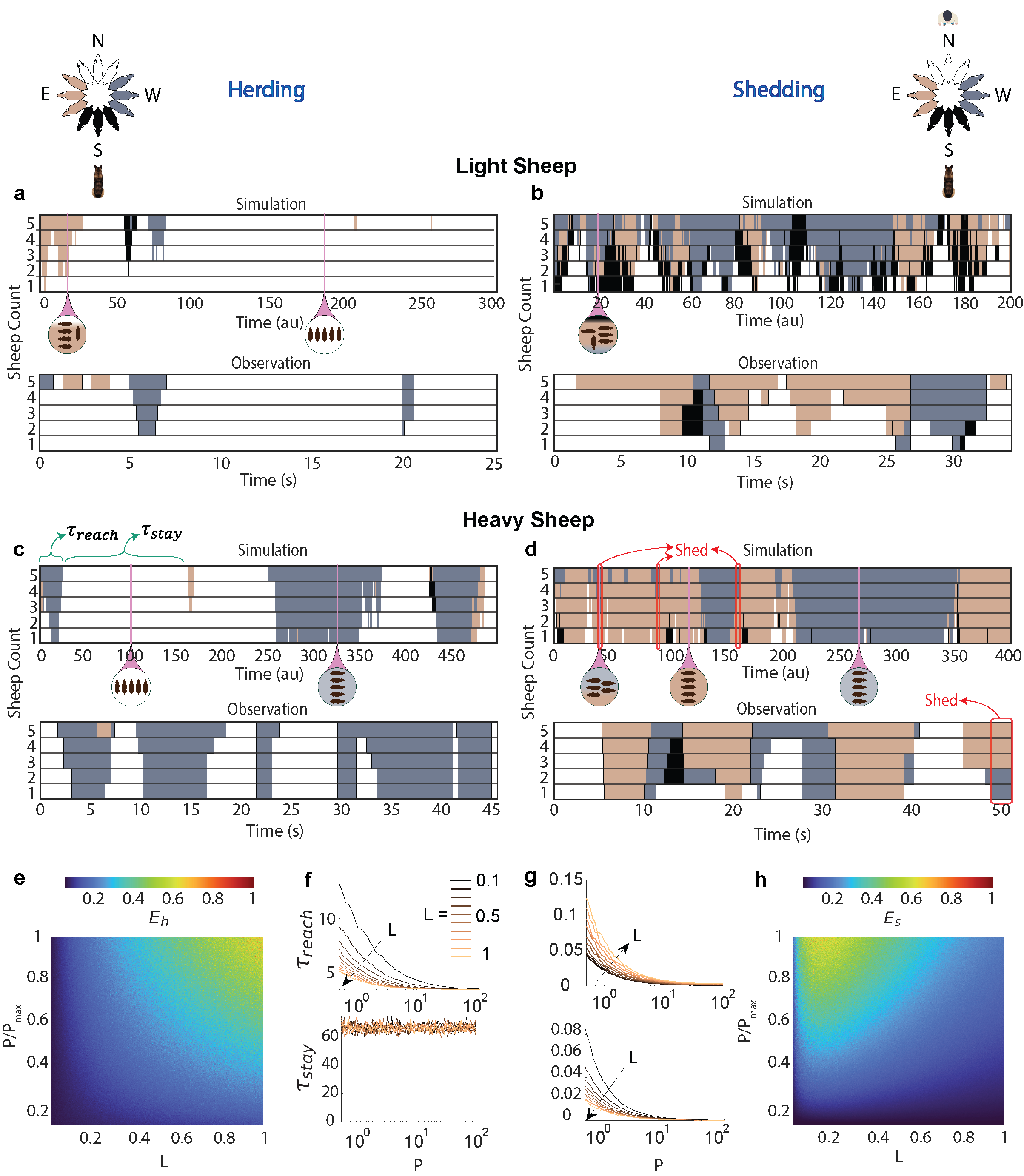}
    \caption{\textbf{Dynamics of Sheep Herding and Shedding.} \textbf{a-d} Time-series of herding and shedding dynamics of a group of 5 sheep with $P = 1$. Starting from a random initial orientation, we plot the time evolution of the number of sheep in each of the four directions. \textbf{a-b} herding and shedding of light sheep $L = 0.9$.   \textbf{Top:} Simulated time-series \textbf{Bottom:} Extracted transition time series from observed videos. Both in simulation and observations, once light sheep reach a herding state (all agents(sheep) in N direction (white)), they remain there with individuals escaping when influenced by noise. For shedding, light sheep panic and individually reorient without any discernible patterns. \textbf{c,d} Herding and shedding process for heavy sheep $L = 0.1$.  \textbf{Top:} Simulated time-series . \textbf{Bottom:} Extracted transition time series from observed videos. Heavy sheep show intermittent herding (all N) followed by synchronous alignment orthogonal to the dog (E or W) by all agents. For shedding, heavy sheep spend a significant time in E and W direction synchronously switching between them. The red vertical lines represent shedding events. Obtaining a shedding event in light sheep is difficult because they panic and randomly reorient when sandwiched between the handler and the dog. However, since heavy sheep synchronously switch between E and W, they provide narrow windows for the dog-handler teams to perform the shed. \textbf{f,g} Effect of pressure and lightness on $\tau_{stay}$ and $\tau_{reach}$ for herding and shedding, respectively. \textbf{e,h} Ease of herding ($E_h$) and ease of shedding ($E_s$) as functions of pressure and lightness. The result shows that it is easier to herd light sheep but easier to shed a group with an intermediate lightness ($L\approx0.1$).}
    \label{fig:Fig3}
\end{figure*}

\subsection*{Modeling Indecisive Sheep Behavior}
To generalize and predict the observed orientation dynamics, we next develop a stochastic model that formalizes the interplay of noise, social interactions, and external stimuli. This framework integrates qualitative insights from empirical observations with quantitative predictions for herding and shedding behaviors.

We model the indecisiveness in sheep behavior during the orientation step using a stochastic framework for  $N_s$ stationary individual agents (sheep). These agents change orientations according to 3  rules:
\begin{enumerate}
    \item \textit{Spontaneous Reorientation}: Agents randomly change direction at a rate $\epsilon$ (noise).
    \item  \textit{Social Influence}: Agents copy the orientation of neighboring agents  at a rate $\gamma$. 
    \item \textit{External Stimulus Response}: Agents reorient in response to external stimuli (dog or handler)  at a rate $\alpha_{ik}$, where $i$ represents the agent's current orientation,  and $k$  represents the position of the stimulus (Figure \ref{fig:Fig2}b).
\end{enumerate}  

These reorientation rules are grounded in empirical observations of sheep-dog trials (see SI Video 10, SI Section 4c). For example, sheep facing the dog (frontal stimulus) panic and reorient randomly  away from it, while those approached from the side reorient to face the direction opposite to the dog’s position. Conversely, sheep facing in the opposite direction from the stimulus remain unaffected (Figure \ref{fig:Fig2}b).


To simplify analysis and exploit rotational symmetry, without  loss of generality, we fix the reference frame such that the external stimulus (dog) is always positioned in the South (S) direction as a convention. To compare our simulation with the experimental data, we bin the orientations of the simulated agents into  4 possible directions: North (N), South (S), East (E), and West (W) (Figure \ref{fig:Fig2}a). 

Traditionally, the dynamics of sheep are modeled with a conventional averaging-based approach, where each sheep averages the orientations of its neighbors and threat from the dog before deciding the direction of motion at each time step (from now on referred to as \textit{averaging agents}) ~\cite{Ranganathan2022-oh,strombom2014solving}. While this fully connected static network approach captures the dynamics of large flocks of sheep, it fails to simulate the indecisive behavior-switching dynamics in small flocks.  Our approach allows agents to be influenced by one factor at a time,  stochastically switching between them. This design is an extension of stochastic choice models that are widely used to capture stochastic switching dynamics in diverse systems across scales ranging from cancer cells and insects to fish schools and human opinion dynamics (such as two state voter models) \cite{sardanyes2018noise, Jhawar2020-uw, Biancalani2014-dr, dyson2015onset, redner2019reality}. For sheep, this method incorporates prior evidence  that small flock  decision-making is Markovian, with frequent switches between leader and follower roles, even in the absence of external stimuli  \cite{pillot2011scalable, gomez2022intermittent}. 

\subsection*{Transition Dynamics and Governing Master Equation}


The dynamics of sheep transitions between orientations can be described  with the following reaction scheme:
\begin{gather}
X_i \xrightarrow{\epsilon} X_{j \neq i} \\
X_i + X_j \xrightarrow{\gamma} 2X_j \\
X_i \xrightarrow{\alpha_{ii}} X_{j \neq i} \quad X_i \xrightarrow{\alpha_{i\tilde i}} \textit{No Effect} \quad
X_i \xrightarrow{\alpha_{ik\neq {i, \tilde i}}} X_{\tilde{k}}.
\end{gather}

Here, \(X_i\) represents a sheep  in the \(i^{\text{th}}\) direction, and \(\alpha_{ik}\) is the influence of a stimulus located in direction $k$ on a sheep in orientation \(X_i\). The notation $\tilde{i}$ indicates the direction opposite to $i$ (e.g., if $i=S$, then $\tilde{i} = N$). Importantly, the total number of sheep remains conserved, such that \(\sum_i X_i = N_s\). 

To capture the time evolution of these transitions, we model the system as a stochastic process governed by a master equation. The master equation is widely used across fields like reaction kinetics, population dynamics, and network theory to describe how probabilities of different system micro-states evolve over time~\cite{erban2009stochastic, bressloff2014stochastic, erban2020stochastic,haag2017modelling}. For sheep orientation dynamics, it captures the interplay of stochastic influences, social interactions, and external stimuli driving collective behaivor.

For this system, the master equation is expressed as:
\begin{align}
\frac{\partial}{\partial t} \mathbf{\mathcal{P}} (\Bar{x}, t) = \sum_{\Bar{x} \neq \Bar{x}'} Tr(\Bar{x} | \Bar{x}') \mathbf{\mathcal{P}}(\Bar{x}') - Tr(\Bar{x}' | \Bar{x}) \mathbf{\mathcal{P}}(\Bar{x}).
\label{Eq:MasterEqn}
\end{align}
Here, \(\Bar{x} = \{x_N,x_S,x_E,x_W\}\) represents the state of the system as the number distribution of sheep oriented in the four orientations, where $x_i$ denotes the number of sheep in the $i^\text{th}$ direction. \(\mathbf{\mathcal{P}}(\Bar{x}, t)\) is the probability of observing  the system in state \(\Bar{x}\) at time \(t\) and $Tr(\Bar{x} | \Bar{x}')$ is rate at which the system transitions from state \(\Bar{x}\) to  \(\Bar{x}'\). 

The master equation captures the rate of change of the probability of observing the system in state \(\Bar{x}\) as the difference between processes bringing the systems into state \(\Bar{x}\) (first term in Equation  (\ref{Eq:MasterEqn}))  and those moving the system away from it (second term in Equation  (\ref{Eq:MasterEqn})). This framework allows us to compute how the collective orientation of sheep evolves over time based on microscopic transition rates, including stochastic influences ($\epsilon$), social interactions ($\gamma$), and external stimuli ($\alpha_{ik}$). Detailed derivations of transition rates $Tr(\Bar{x} | \Bar{y})$ in terms of \(\alpha_{ik}, \gamma, \text{and}  \epsilon \) are provided in SI Section 5. 

\subsection*{Quantifying Pressure and Lightness Metrics}
To bridge the microscopic transition rates of the master equations with the macroscopic behavior of the sheep, we introduce quantitative definitions for pressure and lightness:
\begin{itemize}
    \item \textbf{Pressure} ($P_k= \alpha_{kk} / \gamma$): Quantifies the relative influence of external stimuli (dog/handler) on a sheep facing the stimulus compared to the influence of neighboring sheep. Here, $\alpha_{kk}$ is the  threat imposed by the stimulus on  sheep facing it, while $\gamma$ denotes the influence of other sheep. Physically, $\alpha_{kk}$ can be interpreted as the threat imposed by a stimulus on a sheep looking towards it.
    \item \textbf{Lightness} ($L = \alpha_{jk} / \alpha_{kk}$): Quantifies response isotropy, representing how a sheep oriented perpendicularly (E/W) responds to a stimulus compared to when it is directly facing it (S/N). Subscripts $j$ ad $k$ denote the sheep's orientation and the stimulus's position, respectively ($k = \{S,N\}$ for \{dog, handler\}).   
\end{itemize}

Pressure $P_{k}$ ranges from $0$ to $P_{max}$, where $P_{max}$ is the maximum pressure beyond which sheep begin to move. Since heavy sheep  require a higher pressure to respond,  $P_{max}$ depends on the  sheep's lightness. To compare the dynamics of light and heavy sheep and decouple pressure from lightness, we normalize $P$ by $P_{max}$. 

Lightness $L$ ranges from $0$ to $1$. For ideal light sheep  $(L = 1)$, the dog's threat is isotropic and independent of orientation. In contrast,  ideal heavy sheep $(L = 0)$, only respond to stimuli from the front, ignoring stimuli from other directions (Figure~\ref{fig:Fig2}c). Sheep with intermediate lightness values ($0 < L < 1$) reflect greater responsiveness to frontal stimuli compared to perpendicular stimuli  (Figure~\ref{fig:Fig2}b). 

For simplicity, we assume a linear relationship between lightness $L$ and the influence of the dog $\alpha_{ik}$. Incorporating this allows us to consolidate transition rates as follows:
\begin{itemize}
    \item \(\alpha_{ik} = \alpha\) if \(i = k = \{S, N\}\)
    \item  \(\alpha_{ij} = L \alpha\) if \(i = \{E, W\}\) and \(j = \{S, N\}\) 
    \item \(\alpha_{ij} = 0\) otherwise
\end{itemize}
Irrespective of the lightness, stimuli have no effect on  sheep oriented  opposite to them (\(\alpha_{NS} = \alpha_{SN} = 0\)). 

\subsection*{Dynamics of Herding and Shedding Sheep}
We now use our stochastic model with 2 key parameters ($P$ \& $L$) and Gillespie's algorithm~\cite{asmussen2007stochastic} to simulate sheep dynamics using the master equation (\ref{Eq:MasterEqn}). Our analysis focuses on herding (orient all agents in $N$) and shedding (divide agents into $E$ \& $W$) behaviors for a small group size of $N_s=5$ under constant pressure. Despite its simplicity, the model predicts distinct behaviors for light and heavy sheep in both tasks. 

In herding, light agents ($L = 0.9$) quickly reach a herding state and remain stable, with occasional individual escapes driven by noise \(\epsilon\) (Figure~\ref{fig:Fig3}a Top). Heavy agents ($L = 0.1$), in contrast, exhibit intermittent herding states and frequently align orthogonally to the dog (E or W) in a synchronous manner (Figure~\ref{fig:Fig3}c Top). This behavior mirrors noise-induced switching observed in other small group systems (in the absence of external stimuli), where the reorientation of one individual triggers alignment changes across the group ~\cite{Jhawar2020-uw, Biancalani2014-dr, Buhl2006-dw}. Our model shows that isotropic responses facilitate herding, a result supported by empirical data from sheep-dog trials video (Figure~\ref{fig:Fig3}a,c Bottom, SI Videos 6 and 7). 

In shedding, light agents ($L = 0.9$) frequently reorient without discernible patters due to their isotropic responsiveness (Figure~\ref{fig:Fig3}b Top). Heavy agents ($L = 0.1$), however, align orthogonally to the dog and handler and  synchronously switch between E and W directions due to their selective responsiveness (Figure~\ref{fig:Fig3}d Top), which closely matches the empirical shedding dynamics in  sheep-dog trials (Figure \ref{fig:Fig3}b,d Bottom, SI Videos 8 and 9). Our  model effectively captures nuanced behavioral differences between light and heavy sheep, during herding and shedding, demonstrating strong agreement with  real-world observations (see SI Video 4 and Table S2).

To assess which sheep are easier to control, we quantify herding and shedding success by calculating  reaching time ($\tau_{reach}$) and staying time ($\tau_{stay}$). Reaching time measures how long sheep take to achieve the desired orientation, while staying time indicates how long they remain in that state. Given the time-sensitive nature of sheep-dog trials, we define optimal conditions as those that maximize $\tau_{stay}$ and minimize $\tau_{reach}$. We calculate ease  of herding (or shedding) as $E_{h(s)}=\tau_{\text{stay}} / \tau_{\text{reach}}$. 

Our simulations reveal that in herding,  increasing pressure or lightness reduces $\tau_{reach}$, while $\tau_{stay}$ depends only on noise $\epsilon$ (Figure~\ref{fig:Fig3}f, and SI section 6). These results indicate that dogs use pressure to align  sheep but cannot directly influence how long the alignment persists. The analysis of $E_{h}$ confirms that herding light sheep is easier than herding heavy sheep due to their uniform responsiveness to pressure (Figure~\ref{fig:Fig3}e).

In shedding,  the dog and handler create transient splits within the group, resulting in very short $\tau_{\text{stay}}$. Increasing  pressure reduces $\tau_{reach}$, but higher lightness values lead to longer  $\tau_{reach}$ (Figure~\ref{fig:Fig3}g). The analysis of $E_{s}$ shows that  shedding  very heavy or very light sheep is particularly challenging (Figure~\ref{fig:Fig3}h). In trials,  shedding tasks push the capabilities of the dog-handler team to their limits, as the dog  counteracts the sheep's selfish herding tendencies. The model also predicts that the optimal pressure for both herding and shedding is the maximum stationary pressure $P_{\text{max}}$. Beyond this threshold,  sheep flee uncontrollably (Figure~\ref{fig:Fig3}e,h). In practice, dogs dynamically adjust pressure to account for sheep heterogeneity and changes in lightness, underscoring the complexity of controlling indecisive collectives. 

\begin{figure*}[htbp]
    \centering
    \includegraphics[width = 0.99\textwidth]{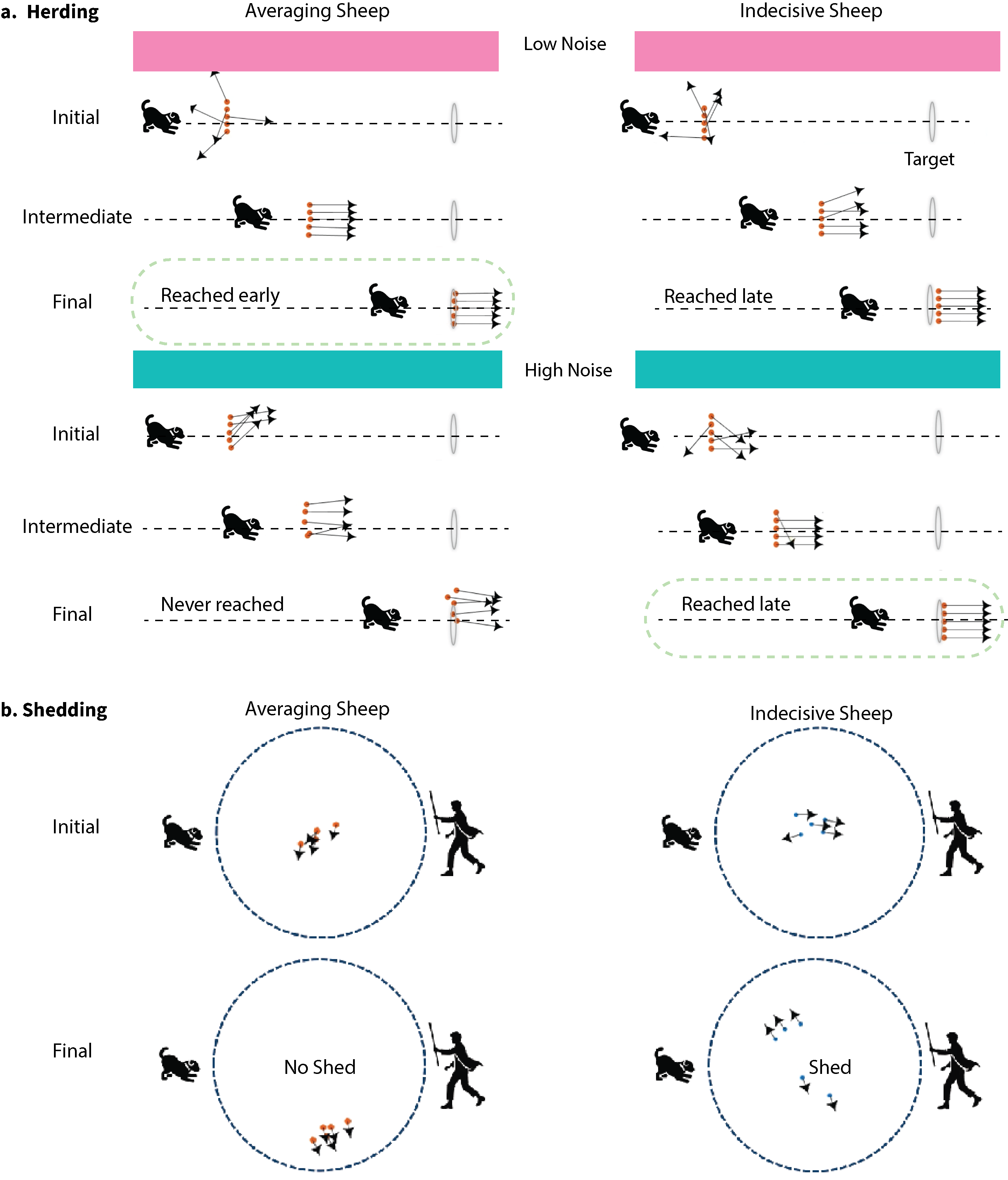}
    \caption{\textbf{Comparison of averaging and indecisive sheep with two-step control}:  \textbf{a} Herding performance is compared under low noise conditions (\(\epsilon/\gamma = 0.08\), shown in magenta, top) and high noise conditions (\(\epsilon/\gamma = 0.8\), shown in teal, bottom). Starting from random initial orientations, both averaging and indecisive agents reach the target under low noise, with averaging agents doing so more quickly. However, under high noise, averaging agents fail to reach the target due to corruption by noise, while indecisive agents, though slower, successfully reach the target. \textbf{b} In shedding tasks, averaging agents fail to split the group, whereas, indecisive agents consistently succeed (SI Video 5) example in figure and SI Video 5: \(\epsilon/\gamma = 0.8\) and \(\alpha/\gamma = 1\)).}
    \label{fig: Fig5}
\end{figure*}

\begin{figure*}[htbp]
    \centering
    \includegraphics[width = 0.99\textwidth]{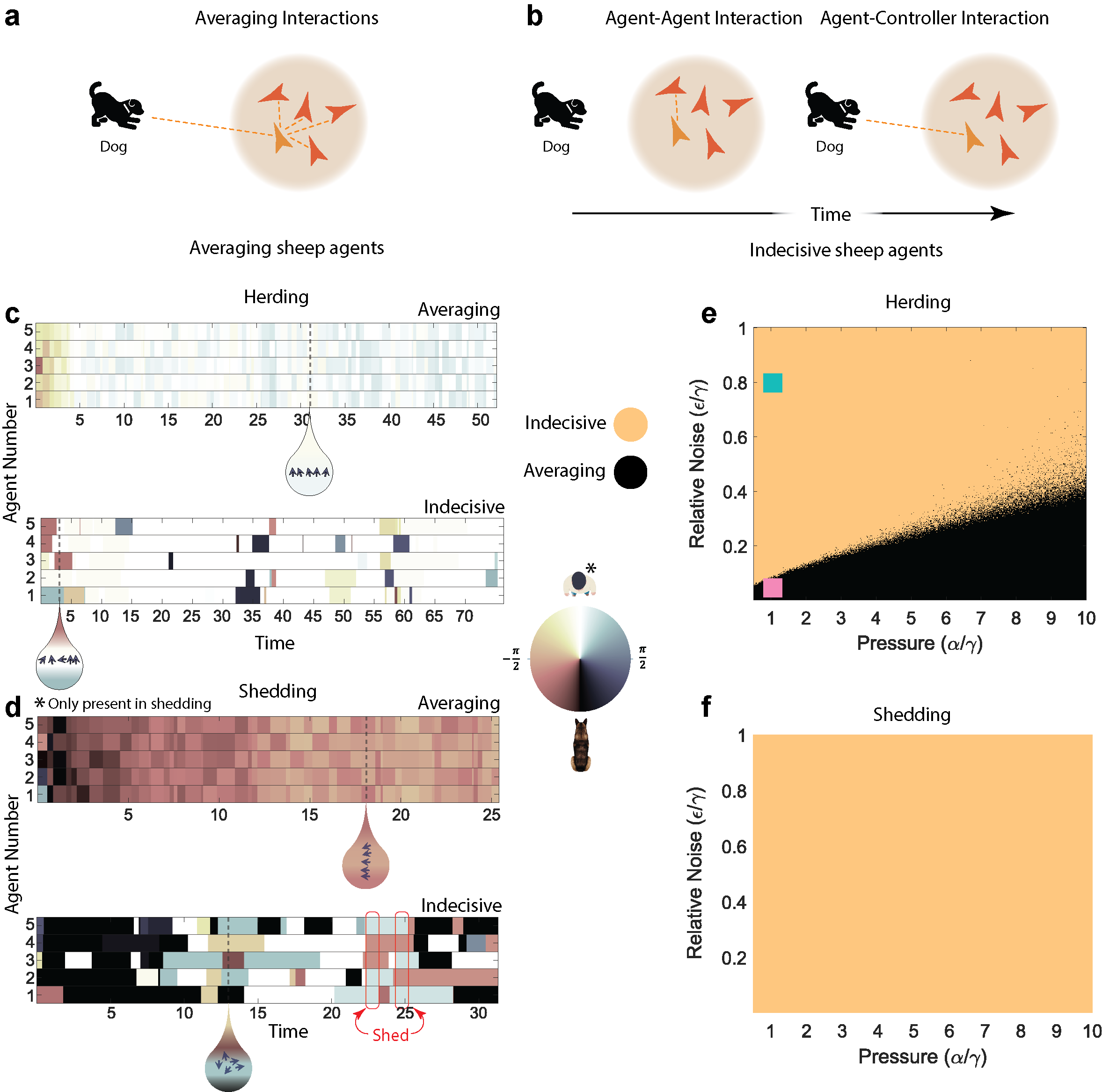}
    \caption{\textbf{Comparison of indecisive and averaging sheep}  \textbf{a,b} Schematic representations of averaging and indecisive sheep agents. Averaging agents average the influence of all factors to update their orientation, while indecisive agents stochastically switch between single influencing factors. \textbf{c-d} Time series showing the dynamics of a small flock of indecisive and averaging sheep during herding and shedding processes for $\alpha = \gamma = \epsilon = 0.1$. \textbf{e,f} Evaluation of ease of herding ($E_h$) and ease of shedding ($E_s$) for both. Averaging sheep are easier to herd under low relative noise, but indecisive sheep are easier to herd in high-noise conditions. For shedding, indecisiveness is crucial since averaging agents fail to split. The teal and magenta squares in \textbf{e} represent the low noise and high noise herding dynamics described in Figure \ref{fig: Fig5}a}
    \label{fig:Fig4}
\end{figure*}

\subsection*{Can Indecisiveness Improve Control?}
Does indecisiveness only pose challenges, or can it aid the dog in controlling the flock?  To investigate this, we next simulate both the \textit{orientation} and \textit{movement} steps of indecisive sheep dynamics in a 2D arena. We extend the 4-direction stochastic framework into continuous 2D space, enabling agents to asynchronously update their orientation to any direction between $-\pi$ to $\pi$, rather than limiting them to discrete directions (N, S, E, W). The rules from Figure~\ref{fig:Fig2}b remain unchanged, but when sheep panic and reorient due to random noise ($\epsilon$) or face the stimulus $\alpha_{kk}$, they randomly select a new direction within the range $-\pi$ to $\pi$, excluding their current orientation. 

We compare these indecisive agents with standard \textit{averaging} agents (Vicsek-type model), where sheep agents synchronously update their position by averaging the effects of all influencing factors (Figure~\ref{fig:Fig4}a,b) \cite{Go2021-sd, Strombom2014-do, Ranganathan2022-oh, Long2019-do}. Both models incorporate alignment with other sheep, repulsion from the dog, and random noise.  For consistency, we focus on ideal light sheep ($L = 1$).

We first simulate the herding problem with the two models (Figure~\ref{fig: Fig5}a). Due to asynchronous updates, indecisive sheep agents move in random directions and diverge, rendering the flock uncontrollable (SI video 5). This result emphasizes the necessity of the two-step control process implemented by shepherd dogs. Integrating this insight into our simulations replicates real dog-herding behaviors, demonstrating that effective control of noisy, indecisive collectives like sheep flocks requires independent regulation of movement and orientation (SI Video 5). 

To compare the controllability of the two models, we simulate the herding process under different noise levels: $\epsilon/\gamma = 0.08$ (Figure~\ref{fig: Fig5}a, top) and $\epsilon/\gamma =0.8$ (Figure~\ref{fig: Fig5}a, bottom). Starting from a random initial orientation,  we find that at low relative noise, averaging agents outperform indecisive agents, reaching the target faster (Figure~\ref{fig: Fig5}a, ``initial" and ``final"). However, at high relative noise, averaging agents fail to reach the target, while indecisive agents, although slower, successfully complete the task. In shedding tasks, averaging agents fail to split into two subgroups under all noise levels, whereas, indecisive agents consistently succeed, regardless of the magnitude of relative noise (Figure~\ref{fig: Fig5}b). 

To quantify control efficiency, we use the metrics ease of herding (shedding), $E_{h (s)}= \tau_{stay}/\tau_{reach}$.  Here, $\tau_{reach}$  is the time required for the flock to achieve the desired orientation (all agents aligned away from the dog in herding, or split into 3 and 2 agents away from the dog and  handler in shedding), and $\tau_{stay}$ is the duration the flock maintains the preferred orientation. 

By analyzing the time series of agent orientations, we observe that in herding, averaging agents maintain orientation but are corrupted by noise, while indecisive agents alternate between epochs of perfect herding and random reorientation (Figure~\ref{fig:Fig4}c). Using $E_h$,  we identify which model is easier to control under  varying pressure ($P = \alpha/\gamma$) and relative noise ($\epsilon/\gamma$). We observe a phase transition:  averaging agents are easier to herd at low noise, but indecisive agents outperform them as noise increases (Figure~\ref{fig:Fig4}e). For shedding, regardless of pressure or noise,  the averaging agents fail to split, whereas  indecisive agents consistently split (Figure~\ref{fig:Fig4}d,f). These results demonstrate that introducing indecisiveness improves control for complex tasks involving both herding and splitting, particularly under noisy conditions.

\subsection*{Developing an Indecisive Swarm Algorithm (ISA)}

Having demonstrated that indecisiveness can improve control, we extend this concept to artificial systems for broader applications. Specifically, we investigate whether stochastic indecisiveness can improve control strategies in robotics, particularly for multi-agent networks navigating constrained trajectories. Control mechanisms that are effective for single agents often fail in multi-agent networked systems due to the emergence of complex global behaviors from simple local interactions between agents. Designing controllers capable of simultaneously managing local, microscopic interactions and global network-level dynamics is still an open problem in control theory \cite{d2023controlling, Piccoli2023-uv}. We show that by introducing stochastic indecisiveness in the multi-agent system, one can enable simple controllers, designed to control single agents, to control multi-agent networked systems.  

To illustrate this, we use the classic trajectory-following problem, a widely studied challenge in robotics \cite{van2024reactive,yildiz2020sliding,punetha2013development}. We model a swarm of agents without visual sensing (blind agents) that rely solely on communicated  orientations from other agents. A controller agent with visual sensing capability is tasked to steer the swarm from an initial position to a target position along a predefined trajectory by applying repulsion forces on individual agents. The problem is a special class of controllability problems called herdability \cite{Ruf2018-wn,
Ruf2018-um,
Shen2025-uo}.  

To evaluate the role of indecisiveness, we develop  the Indecisive Swarm Algorithm (ISA), where agents stochastically switch  inputs between either the controller or another agent Figure~\ref{fig: Fig6}a. ISA agents exhibit two key differences from noisy sheep: (1) they update their dynamics synchronously, and (2) they do not panic in response to the controller or randomly change orientation ($\epsilon = 0$), hence can be programmed to deterministically move away from the controller. Thus, stochastically switching input  sources (controller/agent) remains the sole source of randomness for agents. We benchmark ISA against two standard algorithms: a Vicsek-type Averaging Swarm Algorithm (ASA), where agents average  inputs from all sources \cite{Vicsek1995-gp, Ranganathan2022-oh}, and the Leader-Follower Swarm Algorithm (LFSA), where agents form a fixed hierarchical network, either copying another agent or following the controller non-reciprocally  \cite{jia2019modelling,loria2015leader,cai2023dynamic}.

All three algorithms share two key parameters; i.e., repulsion from the controller \(\tilde{\alpha}\) and alignment with another agent \(\tilde{\gamma}\). The dynamics of the three algorithms are expressed as:

\begin{equation}
   \theta_{ASA}^{(i)}(t) = - \tilde{\alpha} \theta_{ac} + \sum_{agents} \tilde{\gamma} \theta_{aa}
\end{equation}

\begin{equation}
    \theta_{ISA}^{(i)} (t)=
    \begin{cases}
        -\theta_{ac} & \text{with probability } \tilde{\alpha}, \\
        \theta_{aa}  & \text{with probability } \tilde{\gamma}, \\
    \end{cases}
\end{equation}

\begin{equation}
    \theta_{LFSA}^{(i)}(t) =
    \begin{cases}
        -\theta_{ac} & \text{if }  a^{(i)}    \text{ is leader}, \\
        \theta_{aa}  & \text{if } a^{(i)} \text{ is follower}, \\
    \end{cases}
\end{equation}

where
\begin{equation*}
    a^{(i)} =
    \begin{cases}
           \text{leader} & \text{with probability } \tilde{\alpha}, \\
        \text{follower}  & \text{with probability } \tilde{\gamma}. \\
    \end{cases}
\end{equation*}

Here $\theta_{alg.}^{(i)}(t)$ represents the orientation of agent $a_i$ at time $t$ under each algorithm. $\theta_{ac}$ and $\theta_{aa}$ denote the orientation of the agent due to repulsion from the controller and alignment with other agents, respectively. The normalized parameters  $\tilde{\alpha}$ and $\tilde{\gamma}$ are the weights with which  agents respond to the controller and align with others, respectively such that  $\tilde{\alpha} + \tilde{\gamma} = 1$.  We define the stimulus intensity  $I = \tilde{\alpha}/\tilde{\gamma}$, which  generalizes the pressure $(P)$ used in the sheep-dog trials. 

Figure \ref{fig: Fig6}b compares  the trajectories of ISA, ASA, and LFSA for a group size $N_s = 50$. At high stimulus intensity ($I$), all algorithms guide agents along the pre-defined trajectory effectively. However, at low stimulus intensity, ASA and LFSA agents deviate significantly, while ISA agents remain on track, demonstrating the utility of indecisiveness in reducing control effort (SI Video 11).

\subsection*{Temporality and Control Energy}
To better understand why ISA performs better than ASA/LFSA, we examine the  algorithms as special cases of stochastic non-reciprocal temporal networks \cite{li2017fundamental}, where interactions between agents (nodes) evolve dynamically. Such networks are characterized by two timescales: the timescale at which the network restructures $(\tau_{n})$ and the timescale at which  agents update their dynamics $(\tau_{d})$. Agents  update their dynamics every $(\tau_{d})$ by averaging  interactions between consecutive updates (Figure~\ref{fig: Fig6}a). To capture the relationship between these timescales, we define  temporality $(\mathcal{T})$ as:

\begin{gather}
\mathcal{T} =\tau_{d}/\tau_{n}
\end{gather}

This temporality parameter enables us to interpolate between the behaviours of different swarm algorithms (Figure \ref{fig: Fig6}a). When $\mathcal{T} \rightarrow 0$  ($\tau_{n} >> \tau_{d}$), the system mimics LFSA,  where agents interact non-reciprocally in a fixed network topology. When $\mathcal{T} \rightarrow\infty$ ($\tau_{n} << \tau_{d}$), agents average all inputs over time, resembling ASA. At $\mathcal{T} = 1$ ($\tau_{n} = \tau_{d}$), where network restructuring and dynamics updates synchronize, ISA emerges as a distinct behaviour.

Since  we have framed the problem within a temporal network framework, we can apply the concept of control energy, a measure widely used to assess the controllability of complex networks \cite{yan2012controlling, li2017fundamental}. To calculate the control energy required to herd a swarm of agents, we define a safe path as the area  between two boundary lines similar to ref. \cite{van2024reactive}. The controller's task  is to move the swarm to a target position while keeping the swarm's center of mass within the safe path. To prevent  divergence, we also impose bounds on the variance of the swarm (Figure ~\ref{fig: Fig6}c-Top). 

At each dynamics update ($\tau_{d}$), the controller begins with a  low stimulus intensity $I$ and systematically increases $I$ until the swarm moves along the constrained trajectory. For a swarm that reaches the target successfully, we calculate the control energy $\mathcal{E}$ \cite{li2017fundamental} as:

\begin{gather}
\mathcal{E} = \sum_t \frac{1}{2} I_{min}(t)^2
\end{gather}

where $I_{min}(t)$ represents the minimum stimulus intensity required at time $t$ for the swarm to follow the path. The swarm trajectories and the corresponding variation of $I_{min}(t)$ over time are shown in Figure ~\ref{fig: Fig6}c (top and bottom, respectively). 

To evaluate if ISA is the optimal strategy for controlling noisy swarms, we calculate the control energy $\mathcal{E}$ as a function of temporality $\mathcal{T}$ and group size $N_s$. Figure~\ref{fig: Fig6}d shows the variation of $\mathcal{E}$ and the fraction of failed trajectories across different $\mathcal{T}$ values. For $N_s=50$, the failure fraction increases as $\mathcal{T}$ decreases, reflecting the limitations of LFSA. Regardless of group size, the control energy reaches its minimum at $\mathcal{T} = 1$, highlighting the optimality of ISA for herding noisy swarms along predefined paths.

The optimality of ISA can be attributed to its ability to combine the best features of ASA and LFSA. ASA agents quickly align with each other due to their averaging behavior, but they exhibit a strong order that requires the controller to apply high input intensities to reorient the agents in the preferred direction (Figure \ref{fig:Fig6b}a) . For LFSA, due to the network hierarchy, the information from the controller can reach the follower nodes even if the input signal strength is low. However, fixed pairwise non-reciprocal interactions often cause the swarm to split into small clusters, preventing it from reaching the target (Figure \ref{fig:Fig6b}b, SI video 11). ISA reduces the likelihood of cluster formation through network restructuring while maintaining sufficient flexibility to avoid the strong order of ASA agents. As a result, ISA agents successfully reach the target with significantly lower $I_{min}(t)$ compared to ASA (Figure \ref{fig:Fig6b}c).


\begin{figure*}[htbp]
    \centering
    \includegraphics[width = 0.9\textwidth]{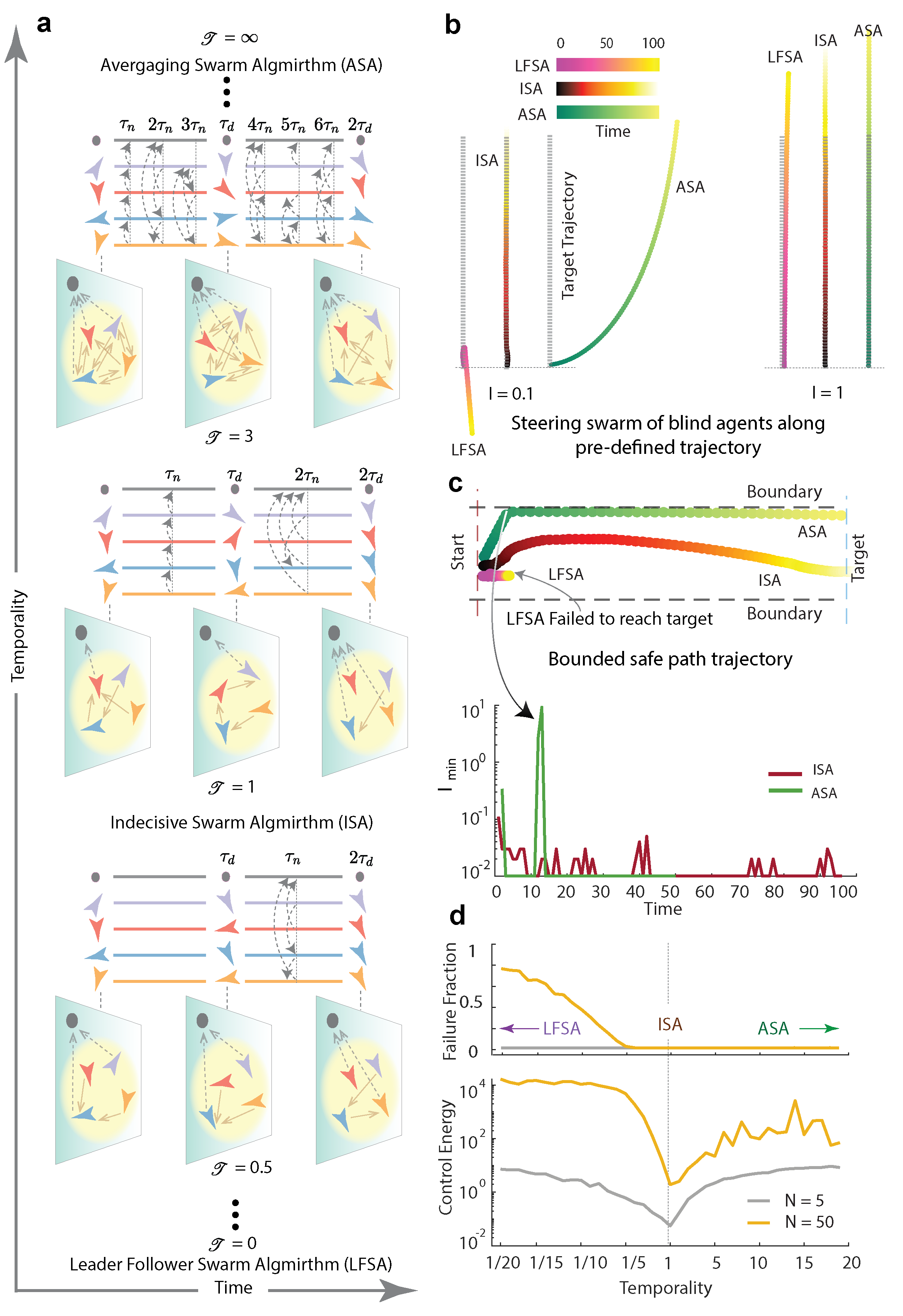}
    \label{fig:Fig5}
\end{figure*}
\begin{figure*}[htbp]
    \centering
    \caption{\textbf{Indecisive Swarm Algorithm.} \textbf{a} A schematic illustrates how temporality $\mathcal{T}$ influences  a swarm of agents  steered by a controller. Each snapshot depicts the  agents' orientation, the controller's relative position, and their interactions. The network restructures every $\tau_n$, and agents update their directions at every dynamic update $\tau_d$ by considering interactions occurring between  consecutive dynamic updates. When $\mathcal{T} = 3$, the  network restructures three times  between two dynamics updates, and  agents calculate a weighted average of all  interactions at each dynamic update to determine their new direction. As $\mathcal{T} \rightarrow \infty$, this behavior resembles an averaging swarm algorithm (ASA), where  agents  average inputs from  all the others and  the controller's repulsion to update their direction. In contrast, when $\mathcal{T} = 0.5$, the network updates  every two dynamic updates. As $\mathcal{T} \rightarrow 0$, the system mimics a  leader-follower swarm algorithm (LFSA), where agents randomly follow a chosen agent or respond  to the controller, indefinitely. At $\mathcal{T} = 1$, the system operates as an indecisive swarm algorithm (ISA), with each network update directly followed by a dynamic update. \textbf{b} Trajectories of LFSA, ISA, and ASA agents are shown for different stimulus intensities $I$ with $N_s=50$. At high stimulus intensity, all agents follow the predefined path. At  low stimulus intensity, only ISA agents stay on the predefined path, while ASA and LFSA agents  deviate significantly. \textbf{c-Top}. The constrained predefined path used for control energy $\mathcal{E}$ calculations.  ISA and ASA agents successfully reach the target, while LFSA agents fail. \textbf{c-Bottom} $I_{min}(t)$ for ISA and ASA agents. The peak in $I_{min}$ for ASA corresponds to the moment when the swarm reaches the boundary. \textbf{d} $\mathcal{E}$ and failure fraction shown  as a functions of $\mathcal{T}$ for 1,500 simulations.  As $\mathcal{T} < 1$, the failure fraction increases for large group sizes ($N_s=50$). Control energy $\mathcal{E}$ achieves a minimum at $\mathcal{T} = 1$, demonstrating  the optimality of ISA for herding noisy agents along predefined trajectories. }
    \label{fig: Fig6}
\end{figure*}

\begin{figure*}[htbp]
    \centering
    \includegraphics[width = 0.99\textwidth]{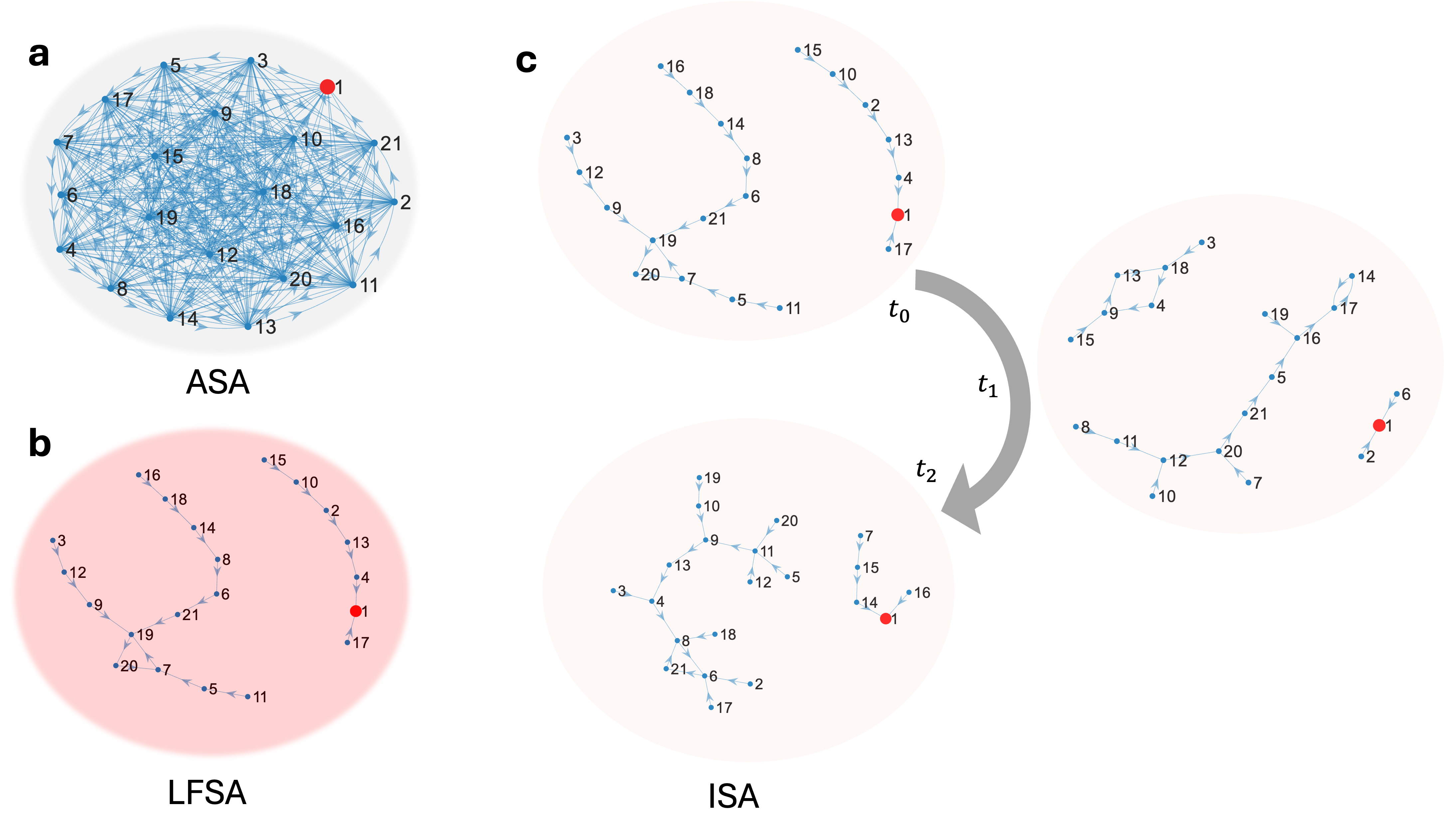}
    \caption{\textbf{ Schematic network representation of ASA, LFSA and ISA.} The red node (node 1) represents the controller. \textbf{a} In ASA, due to the network being fully connected, a high input strength is required from the controller to align the agents in the preferred orientation. \textbf{b} In LFSA, due to the fixed pairwise non-reciprocal interactions, the network often splits in small clusters making it impossible to control. \textbf{c} ISA reduces the likelihood of cluster formation through network restructuring while maintaining sufficient flexibility to avoid the strong order of ASA agents.}
    \label{fig:Fig6b}
\end{figure*}

\section*{Discussion}

\subsection*{Summary}

This study investigates  control mechanisms for noisy, indecisive collectives, using sheepdog trials as a model system. These  trials challenge  trained shepherd dogs to herd and shed (split) small flocks of sheep ($N_s \le 5$), where the dynamics differ markedly from larger flocks. Unlike the cohesive selfish herd behavior seen in large groups, sheep in small flocks  stochastically transition between fleeing (solitary behavior) and following the group (collective behavior),  making them harder to control (i.e., an indecisive herd). By  combining qualitative insights from expert dog handlers  with a  stochastic modeling framework, we analyze  how trained dogs manage these indecisive sheep collectives.  

We find that sheep behavior depends on two key factors:  the dog's threat level and the sheep's switching dynamics. Within the shepherding community, these factors are encapsulated by the terms ``pressure" (the dog's threat) and ``lightness" (the isotropy of the sheep's responsiveness). Light and heavy sheep exhibit distinct behaviors during herding and shedding tasks. To translate this nuanced qualitative knowledge into a quantitative framework, we developed a stochastic model to describe  indecisive  sheep behavior. The model reveals that trained dogs employ a two-step control strategy:  first aligning stationary sheep to a desired orientation (orientation step) before increasing  threat to initiate movement (movement step). Focusing on the orientation step, we modeled sheep as stationary agents that reorient stochastically.  This analysis formalized the concepts of pressure and lightness, confirming their utility as core descriptors of sheep behavior. Comparing the model to data from actual sheepdog trials, we find that high isotropy aids group cohesion (for herding) but complicates splitting, while the dynamics of indecisive sheep are largely governed by the two parameters, pressure and lightness.

We also investigated whether indecisiveness benefits the controller rather than solely posing  a challenge. Extending our framework to  simulate both  orientation and  movement steps in a 2D arena, we compared  indecisive sheep agents with  standard averaging-based Vicsek-type agents. While averaging  agents outperform indecisive agents under low noise conditions, the reverse is true at higher noise levels. For shedding tasks, averaging agents consistently fail to split, while indecisive agents  shed easily, irrespective of noise levels. These results highlight how trained dogs exploit the sheep indecisiveness  as a tool and underscore the importance of the two-step control process.  

Finally, we explored  whether indecisiveness could improve control strategies in artificial systems. Developing the Indecisive Swarm Algorithm (ISA), we compared it against the  Averaging-based Swarm Algorithm (ASA) and Leader-Follower Swarm Algorithm (LFSA) in  a trajectory-following task. ISA agents successfully followed predefined trajectories at low stimulus intensities from the controller, unlike ASA and LFSA agents, which  deviated significantly. Framing swarm algorithms as stochastic temporal networks, we identified two tunable timescales: the dynamics update timescale $(\tau_d)$ and  network  restructuring timescale $(\tau_n)$. By defining  temporality $\mathcal{T}=\tau_d/\tau_n$, we showed that adjusting $\mathcal{T}$ reproduces all three algorithms: ASA  ($\mathcal{T} \to \infty$), LFSA ($\mathcal{T} \to 0$), and ISA ($\mathcal{T} = 1$). Borrowing the concept of control energy from  control theory, we quantified the stimulus intensity required to steer a swarm.  ISA required the least control energy, demonstrating its effectiveness in herding noisy swarms. 

Our findings reveal the counterintuitive advantages of indecisiveness in controlling noisy collectives, with applications ranging from sheepdog trials to artificial swarms. By introducing deliberate indecisiveness, controllers can enhance their ability to perform complex tasks, such as herding and splitting, while also reducing effort in simpler tasks like trajectory-following.



\subsection*{Why Sheepdog Trials are Challenging}

If indecisive agents require less control effort,  why are sheepdog trials considered so challenging? To address this, we extended our indecisive model to large group sizes (SI Section 8). While the model was originally designed to explain the behavior of small groups ($N_s \le 5$) in response to external stimuli, its  extension to larger group sizes captures dynamics consistent with known sheep behaviors. This broader application allowed us to propose a unified phase diagram for indecisive behavior (see SI Section 8 for details), offering insights into transitions between different behavioral regimes as group size and stimulus specificity change.

\begin{figure}[h!t]
    \centering
    \includegraphics[width = 0.5\textwidth]{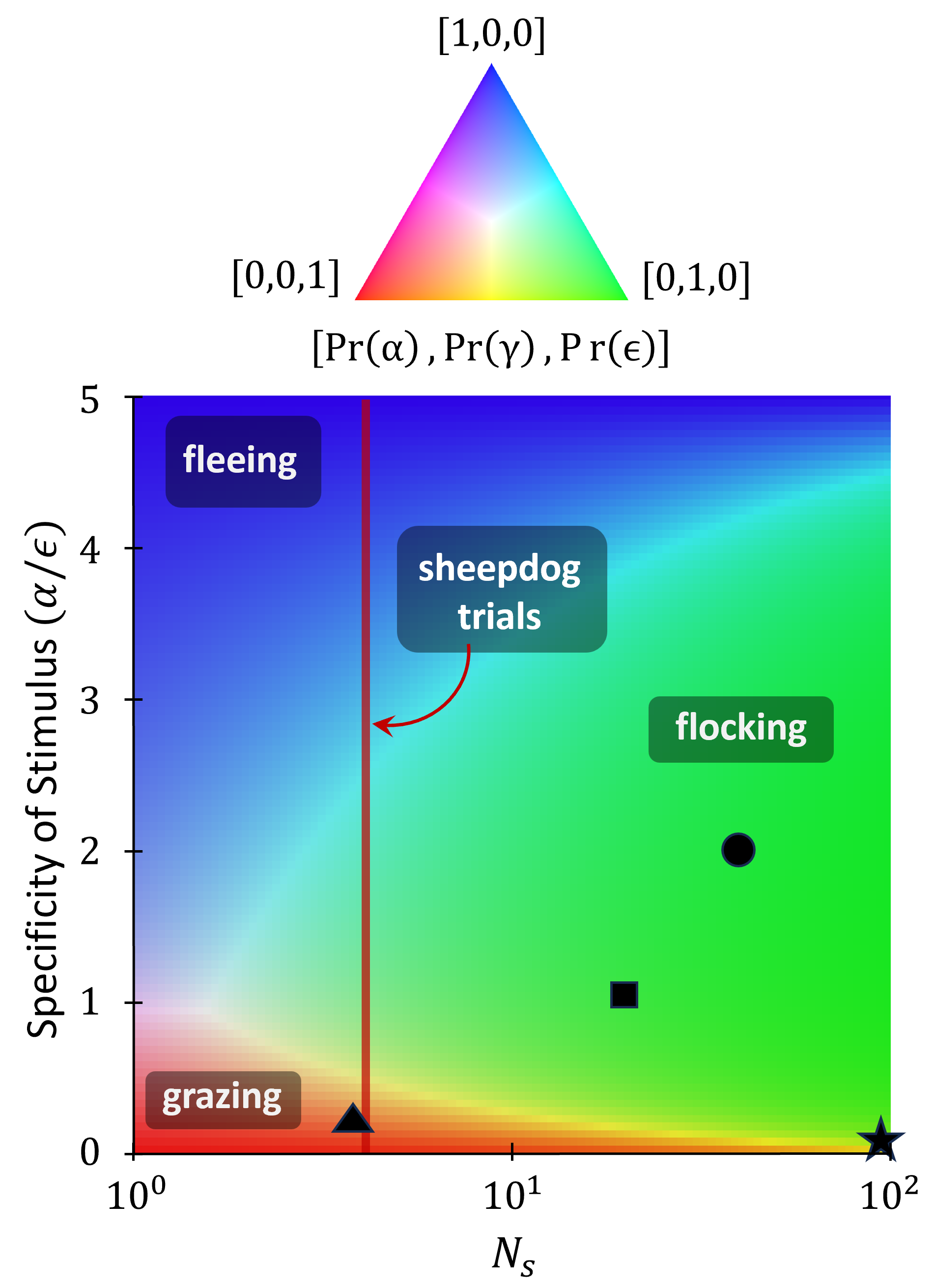}
    \caption{\textbf{Unifying Phase Diagram of Indecisive Collective Behavior.} We present a qualitative phase diagram for indecisive collective behavior as a function of group size ($N_s$) and specificity of external stimulus ($\alpha/\epsilon$) and compare it with the behavior of sheep presented in previous works. We demonstrate that even if indecisive collective model doesn't explicitly explain the selfish herd behavior of sheep in large group sizes, an extension of the model to incorporate group sizes across scales show  show behavior matching the behavior of sheep reported in the literature. Three distinct regimes - fleeing (blue, $\alpha$ dominated), flocking (blue, $\gamma$ dominated), and grazing (green, $\epsilon$ dominated) - are shown. Black shapes represent literature results for different $N_s$ and stimulus specificity ($\alpha/\epsilon$). The $\bullet$ indicates King et. al.'s \cite{King2012-km}  finding that intermediate groups (46 sheep) exhibit selfish herd behavior under threat. The $\blacksquare$ represents  Toulet et. al.'s \cite{toulet2015imitation} study of consensus in intermediate (8 to 32 sheep) following a trained sheep. The $\bigstar$ and $\blacktriangle$ denote  Ginelli et. al.'s \cite{ginelli2015intermittent} and Gomez-Nava et. al.'s \cite{gomez2022intermittent} findings of intermittent flocking and grazing epochs in large (100 sheep) and small groups (4 sheep), respectively. The red line shows the behavior range in sheepdog trials, transitioning from grazing to uncontrolled fleeing through a narrow flocking phase, highlighting the difficulty of managing small indecisive groups of sheep.} 
    \label{fig:Fig7}
\end{figure}

 The phase diagram (Figure~\ref{fig:Fig7}) illustrates the likelihood of individuals being influenced by controlling stimuli ($\alpha$), intra-group interactions ($\gamma$), or random noise (non-specific stimulus) ($\epsilon$). Stimulus specificity, defined as the ratio of  $\alpha/\epsilon$, measures the strength of  external stimuli to noise. External stimuli, such as a dog’s pressure or the departure of an informed sheep, are key factors driving transitions between behaviors.

We identify  three distinct behavioral regimes: flocking (red), dominated by intra-group interaction, resulting in cohesive group behavior; fleeing (blue), dominated by specific stimuli where individuals act independently, ignoring the group;  and grazing (green), dominated by random noise, with individuals disregarding both  specific stimuli and the group. 

In small groups, increasing stimulus specificity shifts behavior from grazing to fleeing. In larger groups,  flocking dominates under typical stimulus intensities. However, when stimulus specificity becomes extremely high - such as during a predator attack  or an encounter with an untrained dog - the flocking phase transitions to fleeing, even in large groups (Figure \ref{fig:Fig7}). 

We validated our model's predictions by comparing them with prior empirical studies of sheep behavior. King et al. \cite{King2012-km} (circle) observed that intermediate-sized groups (46 sheep) exhibited selfish herd behavior under high stimulus specificity,  with herding dogs inducing cohesion. Toulet et al.~\cite{toulet2015imitation} (square) found that when a trained sheep departs  intermediate-sized groups (8-32 sheep), the group reaches a consensus to  follow or ignore the individual, demonstrating the dominance of intra-group interactions  even under mild  stimuli (low  specificity). Ginelli et al.~\cite{ginelli2015intermittent} and Gomez-Nava et al.~\cite{gomez2022intermittent} (star and triangle) studied group dynamics without external stimuli. Ginelli focused on large groups (100 sheep), while Gomez-Nava examined small groups (4 sheep). Both identifyied intermittent  grazing and flocking epochs, aligning with the grazing-flocking transition boundary in our model. These behaviors suggest an evolutionary anticipation of external threats as a defense mechanism.

Our model (red line) predicts that  small groups  transition from grazing to uncontrolled fleeing through a narrow flocking phase as external stimulus increases. This prediction explains why managing small flocks is particularly difficult in sheepdog trials. Since individual sheep vary in their responsiveness to stimuli, effectively herding or splitting small flocks requires the dog to balance  intra-group cohesion with individual responsiveness, as excessive stimulus risks triggering chaotic fleeing. This underscores the complexity of controlling small, indecisive collectives, where behavioral transitions depend on a delicate interplay of external stimuli, noise, and group interactions.


\subsection*{Temporality and Indecisiveness}

Temporal networks have been shown to require significantly less  control energy than  static networks~\cite{li2017fundamental}. This  efficiency arises from their ability to leverage changing topologies to exploit favorable configurations, thereby reducing the need to counteract unfavorable system dynamics. In contrast, static networks, with their fixed structures, often force controllers  to expend substantial energy to navigate energetically costly directions or to overcome inherent system dynamics. A useful analogy is sailing, where  adjusting the sail to align with shifting wind directions enhances efficiency, rather than struggling against them~\cite{li2017fundamental}.

However, this framework assumes  that the controller has prior knowledge of future topology changes. Without such foresight, temporality can actually increase control energy by orders of magnitude compared to static networks~\cite{de2021inherent}. This raises a key question: can temporality still offer advantages in the absence of knowledge about future changes?

We demonstrate that for a specific class of control problems--herding--temporality can significantly reduce  control energy, even without prior knowledge of  topology changes. While traditional  controllability in the context of complex networks involves the ability of the controller in steering the system from any initial state to any desired  state within the state space \cite{d2023controlling, liu2011controllability}, herdability focuses on guiding all agents (nodes) to a fixed consensus state  along a predefined trajectory \cite{Shen2025-uo}.  

Our analysis reveals that indecisive collective--stochastic temporal networks with restructuring timescales equal to system dynamics timescales ($\mathcal{T} = 1$)--are optimal for minimizing control energy. This finding offers a new perspective on leveraging temporality for efficient control of noisy living and robotic swarms, even in the absence of topology foresight.

\subsection*{Broader Implications and Future Directions}

Without external stimuli (e.g., a dog or a handler), our indecisive model extends a general stochastic framework widely applied across diverse systems, including auto-catalytic biochemical reactions~\cite{togashi2001transitions}, heterogeneous cancer cell populations~\cite{sardanyes2018noise}, collective animal movement~\cite{Biancalani2014-dr, Buhl2006-dw, Jhawar2020-uw}, and human opinion dynamics~\cite{caligiuri2023noisy} (SI Section 1, Table S1). By introducing the concept of an external controller, or "shepherd," our analysis establishes a foundational framework for controlling noisy groups in a variety of domains. For instance, Zajdel et al.\cite{zajdel2020scheepdog} demonstrated a shepherd-dog-inspired mechanism to guide cells along specified trajectories, highlighting the potential for shepherding strategies in cellular systems. Building on these insights, our framework could guide the design of effective control mechanisms to herd and sort heterogeneous cell collectives. Such strategies hold  promise for applications like promoting wound healing through coordinated cell movement or selectively isolating healthy cells from infected populations. More broadly, our approach bridges seemingly disparate fields, providing a foundation for  algorithms capable of effectively controlling stochastic, indecisive swarms.

While we presented a simplified model to explore the effects of sheep indecisiveness in sheep-dog-handler interactions,  the real-world dynamics of this system are far more intricate. Shepherd dogs can instinctively predict  sheep movements, but  expertly trained  dogs  uniquely integrate  instinct with handler commands to achieve precise coordination. In successful teams, the handler and dog operate  cohesively, eliminating the need for constant monitoring. Instead, they function as a unified entity, sharing cognitive resources to analyze and anticipate the sheep's behavior in real time \cite{keil2015human, coppinger2002dogs,holland2007herding}. Systematically studying these  interactions, spanning verbal, physical, and visual modalities, could reveal more rich complexities hidden in these multi-species control dynamics, and offering insights into principles of decentralized and stochastic collective control.

\section*{Acknowledgements} We thank the Australian National Sheepdog Trials for allowing us to use the data from their YouTube channel. We also thank Ceri S. Rundle for allowing us to use the data of the sheepdog trials "A way with the dogs" on her YouTube channel CSJ Specialist Canine Feeds. We express our gratitude to Doyle Ivie of Woodsend Stock Dogs for sharing his expertise on sheepdog trials and dog training.  We also thank Matthew Bull, Benjamin Seleb, Atanu Chatterjee, Pankaj Rohilla, and Ishant Tiwari for the discussions and feedback on the manuscript. S.B. acknowledges funding support from NIH MIRA Grant R35GM142588; NSF Grants PHY-2310691; CMMI-2218382; CAREER iOS-1941933; and the Open Philanthropy Project.

\section*{Author contributions} T.C. and S.B. conceptualized the study. T.C. derived the mathematical models, performed the simulations, and analyzed the data. Both the authors contributed to writing the manuscript. S.B. supervised the project.

\section*{Competing interests} The authors declare no competing interests.

\section*{Materials \& Correspondence} The codes are available at https://github.com/bhamla-lab/Controlling-Noisy-Herds-2025

\bibliography{Paperpile_Bib_Controlling_Noisy_Herds}

\begin{thebibliography}{10}
\urlstyle{rm}
\expandafter\ifx\csname url\endcsname\relax
  \def\url#1{\texttt{#1}}\fi
\expandafter\ifx\csname urlprefix\endcsname\relax\def\urlprefix{URL }\fi
\expandafter\ifx\csname doiprefix\endcsname\relax\def\doiprefix{DOI: }\fi
\providecommand{\bibinfo}[2]{#2}
\providecommand{\eprint}[2][]{\url{#2}}

\bibitem{schranz2020swarm}
\bibinfo{author}{Schranz, M.}, \bibinfo{author}{Umlauft, M.}, \bibinfo{author}{Sende, M.} \& \bibinfo{author}{Elmenreich, W.}
\newblock \bibinfo{journal}{\bibinfo{title}{Swarm robotic behaviors and current applications}}.
\newblock {\emph{\JournalTitle{Frontiers in Robotics and AI}}} \textbf{\bibinfo{volume}{7}}, \bibinfo{pages}{36} (\bibinfo{year}{2020}).

\bibitem{sumpter2010collective}
\bibinfo{author}{Sumpter, D.~J.}
\newblock \emph{\bibinfo{title}{Collective animal behavior}} (\bibinfo{publisher}{Princeton University Press}, \bibinfo{year}{2010}).

\bibitem{anderson2019recent}
\bibinfo{author}{Anderson, B.~D.} \& \bibinfo{author}{Ye, M.}
\newblock \bibinfo{journal}{\bibinfo{title}{Recent advances in the modelling and analysis of opinion dynamics on influence networks}}.
\newblock {\emph{\JournalTitle{International Journal of Automation and Computing}}} \textbf{\bibinfo{volume}{16}}, \bibinfo{pages}{129--149} (\bibinfo{year}{2019}).

\bibitem{helbing2001self}
\bibinfo{author}{Helbing, D.}, \bibinfo{author}{Moln{\'a}r, P.}, \bibinfo{author}{Farkas, I.~J.} \& \bibinfo{author}{Bolay, K.}
\newblock \bibinfo{journal}{\bibinfo{title}{Self-organizing pedestrian movement}}.
\newblock {\emph{\JournalTitle{Environment and planning B: planning and design}}} \textbf{\bibinfo{volume}{28}}, \bibinfo{pages}{361--383} (\bibinfo{year}{2001}).

\bibitem{kessels2019traffic}
\bibinfo{author}{Kessels, F.}, \bibinfo{author}{Kessels, R.} \& \bibinfo{author}{Rauscher}.
\newblock \emph{\bibinfo{title}{Traffic flow modelling}} (\bibinfo{publisher}{Springer}, \bibinfo{year}{2019}).

\bibitem{doyle2013feedback}
\bibinfo{author}{Doyle, J.~C.}, \bibinfo{author}{Francis, B.~A.} \& \bibinfo{author}{Tannenbaum, A.~R.}
\newblock \emph{\bibinfo{title}{Feedback control theory}} (\bibinfo{publisher}{Courier Corporation}, \bibinfo{year}{2013}).

\bibitem{lewis2013cooperative}
\bibinfo{author}{Lewis, F.~L.}, \bibinfo{author}{Zhang, H.}, \bibinfo{author}{Hengster-Movric, K.} \& \bibinfo{author}{Das, A.}
\newblock \emph{\bibinfo{title}{Cooperative control of multi-agent systems: optimal and adaptive design approaches}} (\bibinfo{publisher}{Springer Science \& Business Media}, \bibinfo{year}{2013}).

\bibitem{d2023controlling}
\bibinfo{author}{D’Souza, R.~M.}, \bibinfo{author}{di~Bernardo, M.} \& \bibinfo{author}{Liu, Y.-Y.}
\newblock \bibinfo{journal}{\bibinfo{title}{Controlling complex networks with complex nodes}}.
\newblock {\emph{\JournalTitle{Nature Reviews Physics}}} \textbf{\bibinfo{volume}{5}}, \bibinfo{pages}{250--262} (\bibinfo{year}{2023}).

\bibitem{feinerman2018physics}
\bibinfo{author}{Feinerman, O.}, \bibinfo{author}{Pinkoviezky, I.}, \bibinfo{author}{Gelblum, A.}, \bibinfo{author}{Fonio, E.} \& \bibinfo{author}{Gov, N.~S.}
\newblock \bibinfo{journal}{\bibinfo{title}{The physics of cooperative transport in groups of ants}}.
\newblock {\emph{\JournalTitle{Nature Physics}}} \textbf{\bibinfo{volume}{14}}, \bibinfo{pages}{683--693} (\bibinfo{year}{2018}).

\bibitem{Buhl2006-dw}
\bibinfo{author}{Buhl, J.} \emph{et~al.}
\newblock \bibinfo{journal}{\bibinfo{title}{From disorder to order in marching locusts}}.
\newblock {\emph{\JournalTitle{Science}}} \textbf{\bibinfo{volume}{312}}, \bibinfo{pages}{1402--1406} (\bibinfo{year}{2006}).

\bibitem{Jhawar2020-uw}
\bibinfo{author}{Jhawar, J.} \& \bibinfo{author}{Guttal, V.}
\newblock \bibinfo{journal}{\bibinfo{title}{Noise-induced effects in collective dynamics and inferring local interactions from data}}.
\newblock {\emph{\JournalTitle{Philos. Trans. R. Soc. Lond. B Biol. Sci.}}} \textbf{\bibinfo{volume}{375}}, \bibinfo{pages}{20190381} (\bibinfo{year}{2020}).

\bibitem{gomez2022intermittent}
\bibinfo{author}{G{\'o}mez-Nava, L.}, \bibinfo{author}{Bon, R.} \& \bibinfo{author}{Peruani, F.}
\newblock \bibinfo{journal}{\bibinfo{title}{Intermittent collective motion in sheep results from alternating the role of leader and follower}}.
\newblock {\emph{\JournalTitle{Nature Physics}}} \textbf{\bibinfo{volume}{18}}, \bibinfo{pages}{1494--1501} (\bibinfo{year}{2022}).

\bibitem{zha2020opinion}
\bibinfo{author}{Zha, Q.} \emph{et~al.}
\newblock \bibinfo{journal}{\bibinfo{title}{Opinion dynamics in finance and business: a literature review and research opportunities}}.
\newblock {\emph{\JournalTitle{Financial Innovation}}} \textbf{\bibinfo{volume}{6}}, \bibinfo{pages}{1--22} (\bibinfo{year}{2020}).

\bibitem{Koher2019-qo}
\bibinfo{author}{Koher, A.}, \bibinfo{author}{Lentz, H. H.~K.}, \bibinfo{author}{Gleeson, J.~P.} \& \bibinfo{author}{Hövel, P.}
\newblock \bibinfo{journal}{\bibinfo{title}{Contact-based model for epidemic spreading on temporal networks}}.
\newblock {\emph{\JournalTitle{Phys. Rev. X}}} \textbf{\bibinfo{volume}{9}}, \bibinfo{pages}{031017} (\bibinfo{year}{2019}).

\bibitem{masuda2016guide}
\bibinfo{author}{Masuda, N.} \& \bibinfo{author}{Lambiotte, R.}
\newblock \emph{\bibinfo{title}{A guide to temporal networks}} (\bibinfo{publisher}{World Scientific}, \bibinfo{year}{2016}).

\bibitem{humphries2021systematic}
\bibinfo{author}{Humphries, R.}, \bibinfo{author}{Mulchrone, K.}, \bibinfo{author}{Tratalos, J.}, \bibinfo{author}{More, S.~J.} \& \bibinfo{author}{H{\"o}vel, P.}
\newblock \bibinfo{journal}{\bibinfo{title}{A systematic framework of modelling epidemics on temporal networks}}.
\newblock {\emph{\JournalTitle{Applied Network Science}}} \textbf{\bibinfo{volume}{6}}, \bibinfo{pages}{23} (\bibinfo{year}{2021}).

\bibitem{Holme2012-tc}
\bibinfo{author}{Holme, P.} \& \bibinfo{author}{Saramäki, J.}
\newblock \bibinfo{journal}{\bibinfo{title}{Temporal networks}}.
\newblock {\emph{\JournalTitle{Phys. Rep.}}} \textbf{\bibinfo{volume}{519}}, \bibinfo{pages}{97--125} (\bibinfo{year}{2012}).

\bibitem{li2017fundamental}
\bibinfo{author}{Li, A.}, \bibinfo{author}{Cornelius, S.~P.}, \bibinfo{author}{Liu, Y.-Y.}, \bibinfo{author}{Wang, L.} \& \bibinfo{author}{Barab{\'a}si, A.-L.}
\newblock \bibinfo{journal}{\bibinfo{title}{The fundamental advantages of temporal networks}}.
\newblock {\emph{\JournalTitle{Science}}} \textbf{\bibinfo{volume}{358}}, \bibinfo{pages}{1042--1046} (\bibinfo{year}{2017}).

\bibitem{Zhang2021-pa}
\bibinfo{author}{Zhang, X.-Y.}, \bibinfo{author}{Sun, J.} \& \bibinfo{author}{Yan, G.}
\newblock \bibinfo{journal}{\bibinfo{title}{Why temporal networks are more controllable: Link weight variation offers superiority}}.
\newblock {\emph{\JournalTitle{Phys. Rev. Res.}}} \textbf{\bibinfo{volume}{3}} (\bibinfo{year}{2021}).

\bibitem{Lebon2024-kw}
\bibinfo{author}{Lebon, L. C.~G.}, \bibinfo{author}{Lo~Iudice, F.} \& \bibinfo{author}{Altafini, C.}
\newblock \bibinfo{journal}{\bibinfo{title}{On controllability of temporal networks}}.
\newblock {\emph{\JournalTitle{Eur. J. Control}}} \textbf{\bibinfo{volume}{80}}, \bibinfo{pages}{101046} (\bibinfo{year}{2024}).

\bibitem{de2021inherent}
\bibinfo{author}{De~Lellis, P.}, \bibinfo{author}{Di~Meglio, A.}, \bibinfo{author}{Garofalo, F.} \& \bibinfo{author}{Lo~Iudice, F.}
\newblock \bibinfo{journal}{\bibinfo{title}{The inherent uncertainty of temporal networks is a true challenge for control}}.
\newblock {\emph{\JournalTitle{Scientific Reports}}} \textbf{\bibinfo{volume}{11}}, \bibinfo{pages}{6977} (\bibinfo{year}{2021}).

\bibitem{storms2019complex}
\bibinfo{author}{Storms, R.}, \bibinfo{author}{Carere, C.}, \bibinfo{author}{Zoratto, F.} \& \bibinfo{author}{Hemelrijk, C.}
\newblock \bibinfo{journal}{\bibinfo{title}{Complex patterns of collective escape in starling flocks under predation}}.
\newblock {\emph{\JournalTitle{Behavioral ecology and sociobiology}}} \textbf{\bibinfo{volume}{73}}, \bibinfo{pages}{1--10} (\bibinfo{year}{2019}).

\bibitem{brighton2022raptors}
\bibinfo{author}{Brighton, C.~H.} \emph{et~al.}
\newblock \bibinfo{journal}{\bibinfo{title}{Raptors avoid the confusion effect by targeting fixed points in dense aerial prey aggregations}}.
\newblock {\emph{\JournalTitle{Nature Communications}}} \textbf{\bibinfo{volume}{13}}, \bibinfo{pages}{4778} (\bibinfo{year}{2022}).

\bibitem{ruxton2002living}
\bibinfo{author}{Ruxton, G.~D.} \emph{et~al.}
\newblock \emph{\bibinfo{title}{Living in groups}} (\bibinfo{publisher}{Oxford University Press}, \bibinfo{year}{2002}).

\bibitem{noauthor_undated-rn}
\bibinfo{title}{The {Hunt}, {BBC} {Earth}}.
\newblock \bibinfo{howpublished}{https://www.bbcearth.com/shows/the-hunt}.

\bibitem{noauthor_youtube}
\bibinfo{title}{The {Way} {Of} {The} {Cheetah}, {National} {Geographic}}.
\newblock \bibinfo{howpublished}{https://www.youtube.com/watch?v=kTny1iFuTh0}.

\bibitem{sardanyes2018noise}
\bibinfo{author}{Sardany{\'e}s, J.} \& \bibinfo{author}{Alarc{\'o}n, T.}
\newblock \bibinfo{journal}{\bibinfo{title}{Noise-induced bistability in the fate of cancer phenotypic quasispecies: a bit-strings approach}}.
\newblock {\emph{\JournalTitle{Scientific reports}}} \textbf{\bibinfo{volume}{8}}, \bibinfo{pages}{1027} (\bibinfo{year}{2018}).

\bibitem{Biancalani2014-dr}
\bibinfo{author}{Biancalani, T.}, \bibinfo{author}{Dyson, L.} \& \bibinfo{author}{McKane, A.~J.}
\newblock \bibinfo{journal}{\bibinfo{title}{Noise-induced bistable states and their mean switching time in foraging colonies}}.
\newblock {\emph{\JournalTitle{Phys. Rev. Lett.}}} \textbf{\bibinfo{volume}{112}}, \bibinfo{pages}{038101} (\bibinfo{year}{2014}).

\bibitem{dyson2015onset}
\bibinfo{author}{Dyson, L.}, \bibinfo{author}{Yates, C.~A.}, \bibinfo{author}{Buhl, C.} \& \bibinfo{author}{McKane, A.~J.}
\newblock \bibinfo{journal}{\bibinfo{title}{Onset of collective motion in locusts is captured by a minimal model}}.
\newblock {\emph{\JournalTitle{Physical Review E}}} \textbf{\bibinfo{volume}{92}}, \bibinfo{pages}{052708} (\bibinfo{year}{2015}).

\bibitem{redner2019reality}
\bibinfo{author}{Redner, S.}
\newblock \bibinfo{journal}{\bibinfo{title}{Reality-inspired voter models: A mini-review}}.
\newblock {\emph{\JournalTitle{Comptes Rendus Physique}}} \textbf{\bibinfo{volume}{20}}, \bibinfo{pages}{275--292} (\bibinfo{year}{2019}).

\bibitem{grandin2019livestock}
\bibinfo{author}{Grandin, T.}
\newblock \emph{\bibinfo{title}{Livestock handling and transport}} (\bibinfo{publisher}{Cabi}, \bibinfo{year}{2019}).

\bibitem{King2012-km}
\bibinfo{author}{King, A.~J.} \emph{et~al.}
\newblock \bibinfo{journal}{\bibinfo{title}{Selfish-herd behaviour of sheep under threat}}.
\newblock {\emph{\JournalTitle{Curr. Biol.}}} \textbf{\bibinfo{volume}{22}}, \bibinfo{pages}{R561--2} (\bibinfo{year}{2012}).

\bibitem{noauthor_undated-xk}
\bibinfo{title}{History of the {ISDS}}.
\newblock \bibinfo{howpublished}{\url{https://www.isds.org.uk/the-isds/history-of-the-isds/}}.
\newblock \bibinfo{note}{Accessed: 2024-3-16}.

\bibitem{keil2015human}
\bibinfo{author}{Keil, P.~G.}
\newblock \bibinfo{journal}{\bibinfo{title}{Human-sheepdog distributed cognitive systems: An analysis of interspecies cognitive scaffolding in a sheepdog trial}}.
\newblock {\emph{\JournalTitle{Journal of Cognition and Culture}}} \textbf{\bibinfo{volume}{15}}, \bibinfo{pages}{508--529} (\bibinfo{year}{2015}).

\bibitem{holland2007herding}
\bibinfo{author}{Holland, V.~S.}
\newblock \emph{\bibinfo{title}{Herding dogs: Progressive training}} (\bibinfo{publisher}{Turner Publishing Company}, \bibinfo{year}{2007}).

\bibitem{early2020sequential}
\bibinfo{author}{Early, J.}, \bibinfo{author}{Aalders, J.}, \bibinfo{author}{Arnott, E.}, \bibinfo{author}{Wade, C.} \& \bibinfo{author}{McGreevy, P.}
\newblock \bibinfo{journal}{\bibinfo{title}{Sequential analysis of livestock herding dog and sheep interactions}}.
\newblock {\emph{\JournalTitle{Animals}}} \textbf{\bibinfo{volume}{10}}, \bibinfo{pages}{352} (\bibinfo{year}{2020}).

\bibitem{Ranganathan2022-oh}
\bibinfo{author}{Ranganathan, A.}, \bibinfo{author}{Heyde, A.}, \bibinfo{author}{Gupta, A.} \& \bibinfo{author}{Mahadevan, L.}
\newblock \bibinfo{journal}{\bibinfo{title}{Optimal shepherding and transport of a flock}}.
\newblock {\emph{\JournalTitle{arXiv preprint arXiv:2211.04352}}}  (\bibinfo{year}{2022}).

\bibitem{strombom2014solving}
\bibinfo{author}{Str{\"o}mbom, D.} \emph{et~al.}
\newblock \bibinfo{journal}{\bibinfo{title}{Solving the shepherding problem: heuristics for herding autonomous, interacting agents}}.
\newblock {\emph{\JournalTitle{Journal of the royal society interface}}} \textbf{\bibinfo{volume}{11}}, \bibinfo{pages}{20140719} (\bibinfo{year}{2014}).

\bibitem{pillot2011scalable}
\bibinfo{author}{Pillot, M.-H.} \emph{et~al.}
\newblock \bibinfo{journal}{\bibinfo{title}{Scalable rules for coherent group motion in a gregarious vertebrate}}.
\newblock {\emph{\JournalTitle{PloS one}}} \textbf{\bibinfo{volume}{6}}, \bibinfo{pages}{e14487} (\bibinfo{year}{2011}).

\bibitem{erban2009stochastic}
\bibinfo{author}{Erban, R.} \& \bibinfo{author}{Chapman, S.~J.}
\newblock \bibinfo{journal}{\bibinfo{title}{Stochastic modelling of reaction--diffusion processes: algorithms for bimolecular reactions}}.
\newblock {\emph{\JournalTitle{Physical biology}}} \textbf{\bibinfo{volume}{6}}, \bibinfo{pages}{046001} (\bibinfo{year}{2009}).

\bibitem{bressloff2014stochastic}
\bibinfo{author}{Bressloff, P.~C.}
\newblock \emph{\bibinfo{title}{Stochastic processes in cell biology}}, vol.~\bibinfo{volume}{41} (\bibinfo{publisher}{Springer}, \bibinfo{year}{2014}).

\bibitem{erban2020stochastic}
\bibinfo{author}{Erban, R.} \& \bibinfo{author}{Chapman, S.~J.}
\newblock \emph{\bibinfo{title}{Stochastic modelling of reaction--diffusion processes}}, vol.~\bibinfo{volume}{60} (\bibinfo{publisher}{Cambridge University Press}, \bibinfo{year}{2020}).

\bibitem{haag2017modelling}
\bibinfo{author}{Haag, G.}
\newblock \bibinfo{journal}{\bibinfo{title}{Modelling with the master equation}}.
\newblock {\emph{\JournalTitle{Solution Methods}}}  (\bibinfo{year}{2017}).

\bibitem{asmussen2007stochastic}
\bibinfo{author}{Asmussen, S.} \& \bibinfo{author}{Glynn, P.~W.}
\newblock \emph{\bibinfo{title}{Stochastic simulation: algorithms and analysis}}, vol.~\bibinfo{volume}{57} (\bibinfo{publisher}{Springer}, \bibinfo{year}{2007}).

\bibitem{Go2021-sd}
\bibinfo{author}{Go, C.~K.}, \bibinfo{author}{Koganti, N.} \& \bibinfo{author}{Ikeda, K.}
\newblock \bibinfo{title}{Solving the shepherding problem: Imitation learning can acquire the switching algorithm}.
\newblock In \emph{\bibinfo{booktitle}{2021 International Joint Conference on Neural Networks ({IJCNN})}}, \bibinfo{pages}{1--7} (\bibinfo{publisher}{IEEE}, \bibinfo{year}{2021}).

\bibitem{Strombom2014-do}
\bibinfo{author}{Str{\"o}mbom, D.} \emph{et~al.}
\newblock \bibinfo{journal}{\bibinfo{title}{Solving the shepherding problem: heuristics for herding autonomous, interacting agents}}.
\newblock {\emph{\JournalTitle{J. R. Soc. Interface}}} \textbf{\bibinfo{volume}{11}}, \bibinfo{pages}{20140719} (\bibinfo{year}{2014}).

\bibitem{Long2019-do}
\bibinfo{author}{Long, N.~K.}, \bibinfo{author}{Sammut, K.}, \bibinfo{author}{Sgarioto, D.}, \bibinfo{author}{Garratt, M.} \& \bibinfo{author}{Abbass, H.}
\newblock \bibinfo{journal}{\bibinfo{title}{A comprehensive review of shepherding as a bio-inspired {Swarm-Robotics} guidance approach}}.
\newblock {\emph{\JournalTitle{IEEE Transactions on Emerging Topics in Computational Intelligence}}} .

\bibitem{Piccoli2023-uv}
\bibinfo{author}{Piccoli, B.}
\newblock \bibinfo{journal}{\bibinfo{title}{Control of multi-agent systems: results, open problems, and applications}}.
\newblock {\emph{\JournalTitle{arXiv [math.OC]}}}  (\bibinfo{year}{2023}).

\bibitem{van2024reactive}
\bibinfo{author}{Van~Havermaet, S.}, \bibinfo{author}{Khaluf, Y.} \& \bibinfo{author}{Simoens, P.}
\newblock \bibinfo{journal}{\bibinfo{title}{Reactive shepherding along a dynamic path}}.
\newblock {\emph{\JournalTitle{Scientific Reports}}} \textbf{\bibinfo{volume}{14}}, \bibinfo{pages}{14915} (\bibinfo{year}{2024}).

\bibitem{yildiz2020sliding}
\bibinfo{author}{Yildiz, H.}, \bibinfo{author}{Korkmaz~Can, N.}, \bibinfo{author}{Ozguney, O.~C.} \& \bibinfo{author}{Yagiz, N.}
\newblock \bibinfo{journal}{\bibinfo{title}{Sliding mode control of a line following robot}}.
\newblock {\emph{\JournalTitle{Journal of the Brazilian Society of Mechanical Sciences and Engineering}}} \textbf{\bibinfo{volume}{42}}, \bibinfo{pages}{561} (\bibinfo{year}{2020}).

\bibitem{punetha2013development}
\bibinfo{author}{Punetha, D.}, \bibinfo{author}{Kumar, N.}, \bibinfo{author}{Mehta, V.} \emph{et~al.}
\newblock \bibinfo{journal}{\bibinfo{title}{Development and applications of line following robot based health care management system}}.
\newblock {\emph{\JournalTitle{International Journal of Advanced Research in Computer Engineering \& Technology (IJARCET)}}} \textbf{\bibinfo{volume}{2}}, \bibinfo{pages}{2446--2450} (\bibinfo{year}{2013}).

\bibitem{Ruf2018-wn}
\bibinfo{author}{Ruf, S.~F.}, \bibinfo{author}{Egerstedt, M.} \& \bibinfo{author}{Shamma, J.~S.}
\newblock \bibinfo{title}{Herdable systems over signed, directed graphs}.
\newblock In \emph{\bibinfo{booktitle}{2018 Annual American Control Conference (ACC)}}, \bibinfo{pages}{1807--1812} (\bibinfo{publisher}{IEEE}, \bibinfo{year}{2018}).

\bibitem{Ruf2018-um}
\bibinfo{author}{Ruf, S.~F.}, \bibinfo{author}{Egersted, M.} \& \bibinfo{author}{Shamma, J.~S.}
\newblock \bibinfo{journal}{\bibinfo{title}{Herding positive, complex networks}}.
\newblock {\emph{\JournalTitle{arXiv [cs.SY]}}}  (\bibinfo{year}{2018}).

\bibitem{Shen2025-uo}
\bibinfo{author}{Shen, Y.}, \bibinfo{author}{Guan, Y.}, \bibinfo{author}{Xu, S.} \& \bibinfo{author}{Li, Z.}
\newblock \bibinfo{journal}{\bibinfo{title}{Herdability of switching signed networks}}.
\newblock {\emph{\JournalTitle{Automatica (Oxf.)}}} \textbf{\bibinfo{volume}{174}}, \bibinfo{pages}{112124} (\bibinfo{year}{2025}).

\bibitem{Vicsek1995-gp}
\bibinfo{author}{Vicsek, T.}, \bibinfo{author}{Czir{\'o}k, A.}, \bibinfo{author}{Ben-Jacob, E.}, \bibinfo{author}{Cohen, I., I} \& \bibinfo{author}{Shochet, O.}
\newblock \bibinfo{journal}{\bibinfo{title}{Novel type of phase transition in a system of self-driven particles}}.
\newblock {\emph{\JournalTitle{Phys. Rev. Lett.}}} \textbf{\bibinfo{volume}{75}}, \bibinfo{pages}{1226--1229} (\bibinfo{year}{1995}).

\bibitem{jia2019modelling}
\bibinfo{author}{Jia, Y.} \& \bibinfo{author}{Vicsek, T.}
\newblock \bibinfo{journal}{\bibinfo{title}{Modelling hierarchical flocking}}.
\newblock {\emph{\JournalTitle{New Journal of Physics}}} \textbf{\bibinfo{volume}{21}}, \bibinfo{pages}{093048} (\bibinfo{year}{2019}).

\bibitem{loria2015leader}
\bibinfo{author}{Loria, A.}, \bibinfo{author}{Dasdemir, J.} \& \bibinfo{author}{Jarquin, N.~A.}
\newblock \bibinfo{journal}{\bibinfo{title}{Leader--follower formation and tracking control of mobile robots along straight paths}}.
\newblock {\emph{\JournalTitle{IEEE transactions on control systems technology}}} \textbf{\bibinfo{volume}{24}}, \bibinfo{pages}{727--732} (\bibinfo{year}{2015}).

\bibitem{cai2023dynamic}
\bibinfo{author}{Cai, H.}, \bibinfo{author}{Guo, S.} \& \bibinfo{author}{Gao, H.}
\newblock \bibinfo{journal}{\bibinfo{title}{A dynamic leader--follower approach for line marching of swarm robots}}.
\newblock {\emph{\JournalTitle{Unmanned Systems}}} \textbf{\bibinfo{volume}{11}}, \bibinfo{pages}{67--82} (\bibinfo{year}{2023}).

\bibitem{yan2012controlling}
\bibinfo{author}{Yan, G.}, \bibinfo{author}{Ren, J.}, \bibinfo{author}{Lai, Y.-C.}, \bibinfo{author}{Lai, C.-H.} \& \bibinfo{author}{Li, B.}
\newblock \bibinfo{journal}{\bibinfo{title}{Controlling complex networks: How much energy is needed?}}
\newblock {\emph{\JournalTitle{Physical review letters}}} \textbf{\bibinfo{volume}{108}}, \bibinfo{pages}{218703} (\bibinfo{year}{2012}).

\bibitem{toulet2015imitation}
\bibinfo{author}{Toulet, S.}, \bibinfo{author}{Gautrais, J.}, \bibinfo{author}{Bon, R.} \& \bibinfo{author}{Peruani, F.}
\newblock \bibinfo{journal}{\bibinfo{title}{Imitation combined with a characteristic stimulus duration results in robust collective decision-making}}.
\newblock {\emph{\JournalTitle{PloS one}}} \textbf{\bibinfo{volume}{10}}, \bibinfo{pages}{e0140188} (\bibinfo{year}{2015}).

\bibitem{ginelli2015intermittent}
\bibinfo{author}{Ginelli, F.} \emph{et~al.}
\newblock \bibinfo{journal}{\bibinfo{title}{Intermittent collective dynamics emerge from conflicting imperatives in sheep herds}}.
\newblock {\emph{\JournalTitle{Proceedings of the National Academy of Sciences}}} \textbf{\bibinfo{volume}{112}}, \bibinfo{pages}{12729--12734} (\bibinfo{year}{2015}).

\bibitem{liu2011controllability}
\bibinfo{author}{Liu, Y.-Y.}, \bibinfo{author}{Slotine, J.-J.} \& \bibinfo{author}{Barab{\'a}si, A.-L.}
\newblock \bibinfo{journal}{\bibinfo{title}{Controllability of complex networks}}.
\newblock {\emph{\JournalTitle{nature}}} \textbf{\bibinfo{volume}{473}}, \bibinfo{pages}{167--173} (\bibinfo{year}{2011}).

\bibitem{togashi2001transitions}
\bibinfo{author}{Togashi, Y.} \& \bibinfo{author}{Kaneko, K.}
\newblock \bibinfo{journal}{\bibinfo{title}{Transitions induced by the discreteness of molecules in a small autocatalytic system}}.
\newblock {\emph{\JournalTitle{Physical review letters}}} \textbf{\bibinfo{volume}{86}}, \bibinfo{pages}{2459} (\bibinfo{year}{2001}).

\bibitem{caligiuri2023noisy}
\bibinfo{author}{Caligiuri, A.} \& \bibinfo{author}{Galla, T.}
\newblock \bibinfo{journal}{\bibinfo{title}{Noisy voter models in switching environments}}.
\newblock {\emph{\JournalTitle{Physical Review E}}} \textbf{\bibinfo{volume}{108}}, \bibinfo{pages}{044301} (\bibinfo{year}{2023}).

\bibitem{zajdel2020scheepdog}
\bibinfo{author}{Zajdel, T.~J.}, \bibinfo{author}{Shim, G.}, \bibinfo{author}{Wang, L.}, \bibinfo{author}{Rossello-Martinez, A.} \& \bibinfo{author}{Cohen, D.~J.}
\newblock \bibinfo{journal}{\bibinfo{title}{Scheepdog: programming electric cues to dynamically herd large-scale cell migration}}.
\newblock {\emph{\JournalTitle{Cell systems}}} \textbf{\bibinfo{volume}{10}}, \bibinfo{pages}{506--514} (\bibinfo{year}{2020}).

\bibitem{coppinger2002dogs}
\bibinfo{author}{Coppinger, R.} \& \bibinfo{author}{Coppinger, L.}
\newblock \emph{\bibinfo{title}{Dogs: a new understanding of canine origin, behavior and evolution}} (\bibinfo{publisher}{University of Chicago Press}, \bibinfo{year}{2002}).

\end{thebibliography}

\end{document}


\maketitle

\section{Behavior switching across scales}
From simple organisms like ants to complex ones like humans can switch between different behaviors depending on group size, environmental factors, or just because of noise. The indecisive model presented in the main text is a special case of such switching dynamics, where sheep in small flocks, in the presence of a shepherd dog, switch between following other sheep and trying to flee.  Here, we list the switching behavior in different organisms reported in the literature to demonstrate that such switching dynamics are common in organisms across scales.

\begin{table}[h!]
\centering
\caption*{Table S1: Switching dynamics in animal groups reported in the literature}
\resizebox{\textwidth}{!}{%
\begin{tabular}{|c|c|c|}
\hline
System & Description & Reference \\ \hline

Ants   & Carrier ants switch between
two possible roles, pullers and lifters.   & \cite{feinerman2018physics}  \\ \hline

Locusts   & Switching dynamics between driving and hiding when attacked by predator   & \cite{Buhl2006-dw}  \\ \hline

Golden shiners fish & Switch between swarming, schooling and milling & \cite{tunstrom2013collective}  \\ \hline

Cichlid fish & Switching between following other fish and randomly choosing a direction and moving   & \cite{Jhawar2020-uw}   \\ \hline

Seals   & Switch between different interaction rules depending on presence or absence of predatory risk   & \cite{de2013movement}  \\ \hline

Pigeons & Switching between aligning with neighbor and turning away from the predator   & \cite{sankey2021absence}   \\ \hline

Starlings & Switching complex dynamical patterns when escaping from predator   & \cite{storms2019complex}   \\ \hline

Cliff swallows & Switch between relaxing and being alert and the ratio of time spent between the two increases with group size  & \cite{brown1987group} \\ \hline

Sheep  & Switching between being leader and being follower  & \cite{gomez2022intermittent}  \\ \hline

Humans & Switching between opinions & \cite{zha2020opinion}  \\ \hline

\end{tabular}
}
\label{Table:S1}
\end{table}

\section{History of Sheepdog Trials}

Herding animals with the aid of dogs traces its roots far into history. Archaeological findings indicate that dogs were among the first animals to be domesticated by humans, with evidence of domesticated dogs existing over 15,000 years ago \cite{ahmad2020domestication}. Initially, these early domesticated dogs primarily assisted in hunting. The earliest depictions of dog-assisted hunting can be found in rock art from sites such as Shuwaymis and Jubbah in northwestern Saudi Arabia, dating back to 7000 BCE (Figure 1a) \cite{guagnin2018pre}.

While evidence of domesticated sheep and goats dates back to 8500 BCE \cite{alberto2018convergent}, the oldest remains of both domesticated dogs and sheep discovered together in archaeological excavations date back to 1735 BCE \cite{olsen1985prehistoric}. During the Bronze Age, shepherd dogs emerged as guarding dogs, protecting animals from various wild predators (Figure 1b). However, these dogs had limited interaction with the animals. Shepherd dogs have also been a part of early Greek and Roman culture and mythology (Figure 1c) \cite{Dogs_Ancient,WaltersArt}. References to sheepdogs can be found in the writings of Cato the Elder and Marcus Terentius Varro dating back to 100-200 BC \cite{coppinger2000dogs}, as well as in the Old Testament (Job 30:1).

The evolution of modern herding dogs took shape many centuries after the emergence of livestock guardian dogs. Herding breeds were well-established throughout Europe by the 17th century \cite{hancock2014dogs}, Interestingly, the border-collie breed originated with Old Hemp, a notable stud dog owned by Adam Telfer in the late 19th century (Figure 1d) \cite{halsall1980sheepdogs}.

The competition sheepdog trials began in the 1870s in Wales and served as gatherings where shepherds could showcase and compare the skills of their dogs. The first recorded sheepdog trial occurred in 1873 near Bala, Wales, where shepherds competed to determine the most adept border collies. The first international sheepdog trials competition was held at Alexandra Palace in London in June 1876 (Figure 1e) \cite{Border_Collie}.]

\begin{figure*}[t]
    \centering
    \includegraphics[width = .99\textwidth]{Figure2_SI.png}
    \caption{History of herding dogs: \textbf{a.} The oldest cave painting of domesticated dogs dating back to 7000 BCE found in Shuwaymis and Jubbah in northwestern Saudi Arabia \cite{guagnin2018pre}. \textbf{b.} Dog-assisted herding in Norway from the Bronze Age. \textbf{c.} Roman relief titled 'Relief of a Herdsman' at the Walters Art Museum portraying the Greek myth of Endymion, a handsome shepherd, his dog, a goat, and a sheep \cite{WaltersArt}. \textbf{d.} Photo of Old Hemp, a stud dog known to be the father of modern Border Collie \cite{halsall1980sheepdogs} \textbf{e.} Photo of the first international sheepdog trials at Alexandra Palace in London in 1876 \cite{Border_Collie}}
     
    \label{fig:SI2}
    \end{figure*} 



\section{Details of modern sheepdog trials competitions}

In modern sheep-dog trial competitions, the contesting dog handlers bring their trained dogs to an arena that is specifically designed for the competition. Each dog-handler duo is presented with a randomly selected group of sheep, typically consisting of 3 to 5 individuals. There are various types of tasks that each team has to perform in a given amount of time. A panels of judges score the dog on its ability to skillfully maneuver the flock of sheep. 

Standard competitions generally have 5 tasks in the order of \textit{fetch, drive, shed, pen, and single}.  The initial set-up for the trials consists of a handler post and sheep grazing far from the post (Figure 2a). The handler must stand at the post for the entire fetch and drive; failing to do so would result in a deduction in points. The trial starts with a \textit{fetch} (Figure 2b), where the dog has to bring the flock of sheep to the handler post in a straight line and make a tight turn around the handler post to start the first step of the \textit{drive}. In the \textit{drive} the dog has to move the sheep away from the handler and bring them to the shedding ring (Figure 2c-d). Once the sheep are in the shedding ring, the handler is allowed to leave the post and the team performs a \textit{shed} where they split 2 sheep from the flock (Figure 2e). After the shed, the team has to move the sheep inside a small gated arena called the \textit{pen} (Figure 2f). The \textit{penned} sheep then are taken out to the shedding ring for one last time to perform a \textit{single} where one sheep is separated from the flock and that ends the trial (Figure 2g). The teams need to complete the tasks within a stipulated time. Tasks not completed within the time get zero points. For each mistake, the judges deduct points and the team with the maximum points wins the competition. 

\begin{figure*}[t]
    \centering
    \includegraphics[width = .9\textwidth]{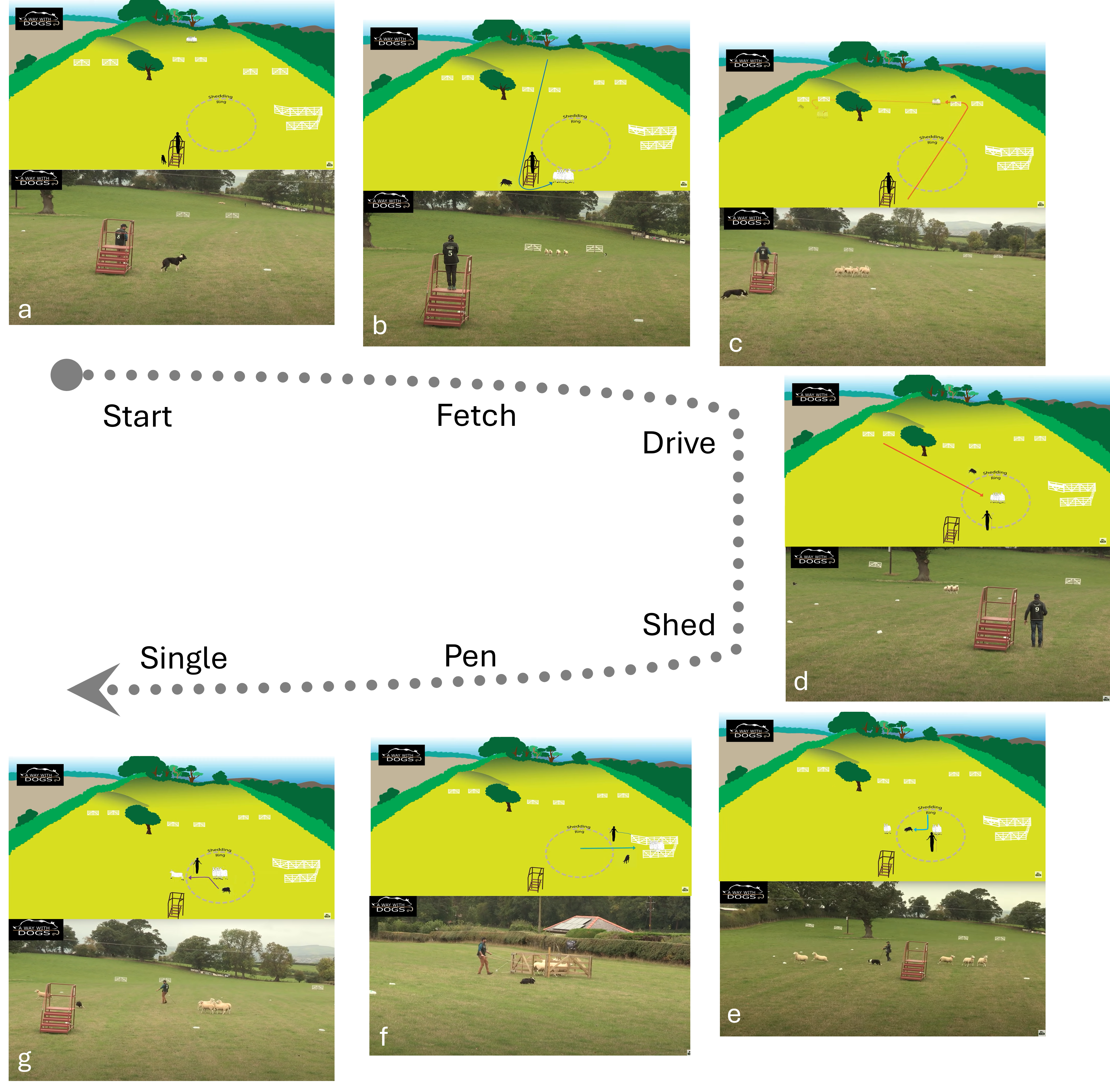}
    \caption{Sheepdog trials competition and task sequence}
    \label{fig:SI1}
    \end{figure*}


\section{Data analysis and 4-direction model}

\subsection*{a. Data analysis}
Traditionally, top-view videos of the herding process are collected by a drone and analyzed to track the trajectories of the sheep and the dog ~\cite{gomez2022intermittent,jadhav2024collective}. However, we were informed that in sheepdog trial competitions, the dog might get distracted by the drone and therefore, it was not possible to collect top-view videos of the actual competition. Fortunately, the owners of the YouTube Channels CSJ Specialist Canine Feeds and Australian National Sheepdog Trial allowed us to use all the recorded videos on their channel for research. This enabled us to use 25+ hours of sheepdog trial competitions with expert commentary to gather a lot of qualitative information by comparing the commentary with the handler-dog-sheep dynamics. However, due to the side-view cameras used in the videos, we could only manually quantify data with 4-direction resolution, i.e., towards the dog, away from the dog, perpendicular left to the dog, and perpendicular right to the dog. Since we calculate the orientation of the sheep with respect to the dog, without loss of generality, we fix the dog's position towards the South (S) direction and calculate the orientation of the sheep accordingly.

\subsection*{b. The two-step process}
We observe that the dog implements a two-step process to control sheep (See SI video 2). To quantify the success of trained dogs in implementing the process, we manually note the movement and halting of the sheep flocks. Ideally, the stationary flock should begin to move only when all the sheep are oriented in the preferred (N) direction. Conversely, if any sheep in a flock moving in N direction changes its orientation, the flock must halt. To quantify the dog's success in achieving this process, we analyzed 106 instances of stationary flocks and 119 instances of moving flocks. We find that out of 106 instances of stationary flocks, in 87 instances, the sheep only move when they are all in the preferred (N) orientation i.e., opposite to the dog's position (S) (Figure 3a). Similarly, out of 119 instances of moving flock in the N direction, in 99 instances, the flock stops when one or more sheep turn away from the preferred (N) direction (i.e., E, W, or S) (Figure 3b). Although the exact mechanism of how the dog achieves this is beyond the scope of the paper, our analysis demonstrates that trained dogs can successfully implement two-step process.

\begin{figure*}[htbp]
    \centering
    \includegraphics[width = .9\textwidth]{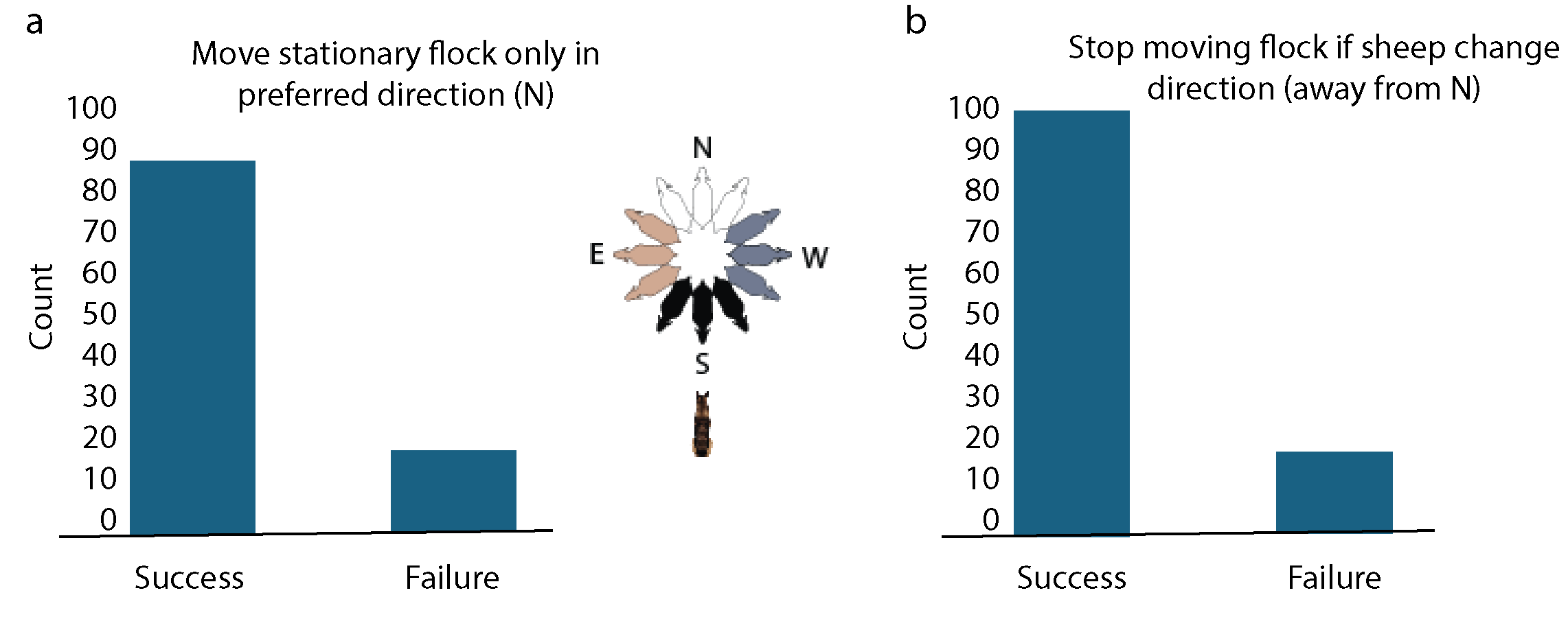}
    \caption{Two-step process: We manually quantified the success of the dog in implementing the two-step process. \textbf{a} Out of 106 instances of stationary flocks, we observe that in 87 instances the dog successfully move the flock only when all the sheep are oriented in the preferred (N) direction. \textbf{b} Out of 119 instances of moving flocks, we observe that in 99 instances the dog was successful in halting the movement when one or more sheep turn in the direction away from N (\textit{Videos Used in the analysis: 2024 National Sheep Dog Trial Championship series and 2023 National Sheep Dog Trial Championship series in YouTube channel Australian National Sheepdog Trial and A Way with the dogs 2 series on YouTube channel CSJ Specialist Canine Feeds. In this analysis, we found some sheep are extremely panicky and bolt (See https://www.youtube.com/watch?v=hqd9EjPTbtM\&list=PLlJ1Ji8GrEqkBir3xiOe-m2PpakJlzoIL\&index=3). Those flocks are considered impossible to work with. We exclude those videos from our analysis.}}
    \label{fig:SI3}
    \end{figure*}

\subsection*{c. Transition rules}
To compare the results of the model with the observed data, we developed a 4-direction stochastic model. The rules of the transition of the sheep (Figure 2b of main text) are motivated by the observed interactions of the sheep. For instance, we observe that stationary sheep can change orientation when influenced by a dog or copy the orientation of other sheep. In many scenarios, it is impossible to distinguish between the two influences and in some cases, the resulting final orientational state could be a result of both the influences applied sequentially. Furthermore, there are certain instances, where a sheep can randomly switch to a direction without any apparent reason. Therefore, in our model, we choose three interaction rules-- influence by the dog, influence by other sheep, and influence by random noise. 

Further, we observe that the influence of the dog results in significantly different outputs depending on the initial orientation of the sheep. Sheep oriented towards the dog (S) tend to randomly reorient in any direction (E, W or N) (Figure 4a-b) while sheep oriented perpendicular to the dog (E or W) reorients in the opposite direction of the dog (N) (Figure 4c-d). To quantify the effect, we looked at 80 flocks of stationary sheep. As the dog increases the pressure by moving towards the flock, we only note the first orientational transition in the flock since the subsequent transitions could be a sheep-sheep alignment rather than the influence of the dog. We exclude instances of milling where sheep start rotating clockwise or counterclockwise at the same position. We found that when a sheep is initially oriented towards the dog (S) (Figure 4e), it reorients to directions N, E and W with similar probabilities (Figure 4g). However, when a sheep is oriented perpendicular to the dog E or W (Figure 4f), the chances of them orienting to N is significantly higher (Figure 4h). This result motivated the rules for our model defined in Figure 2b in the main text.

\begin{figure*}[htbp]
    \centering
    \includegraphics[width = .99\textwidth]{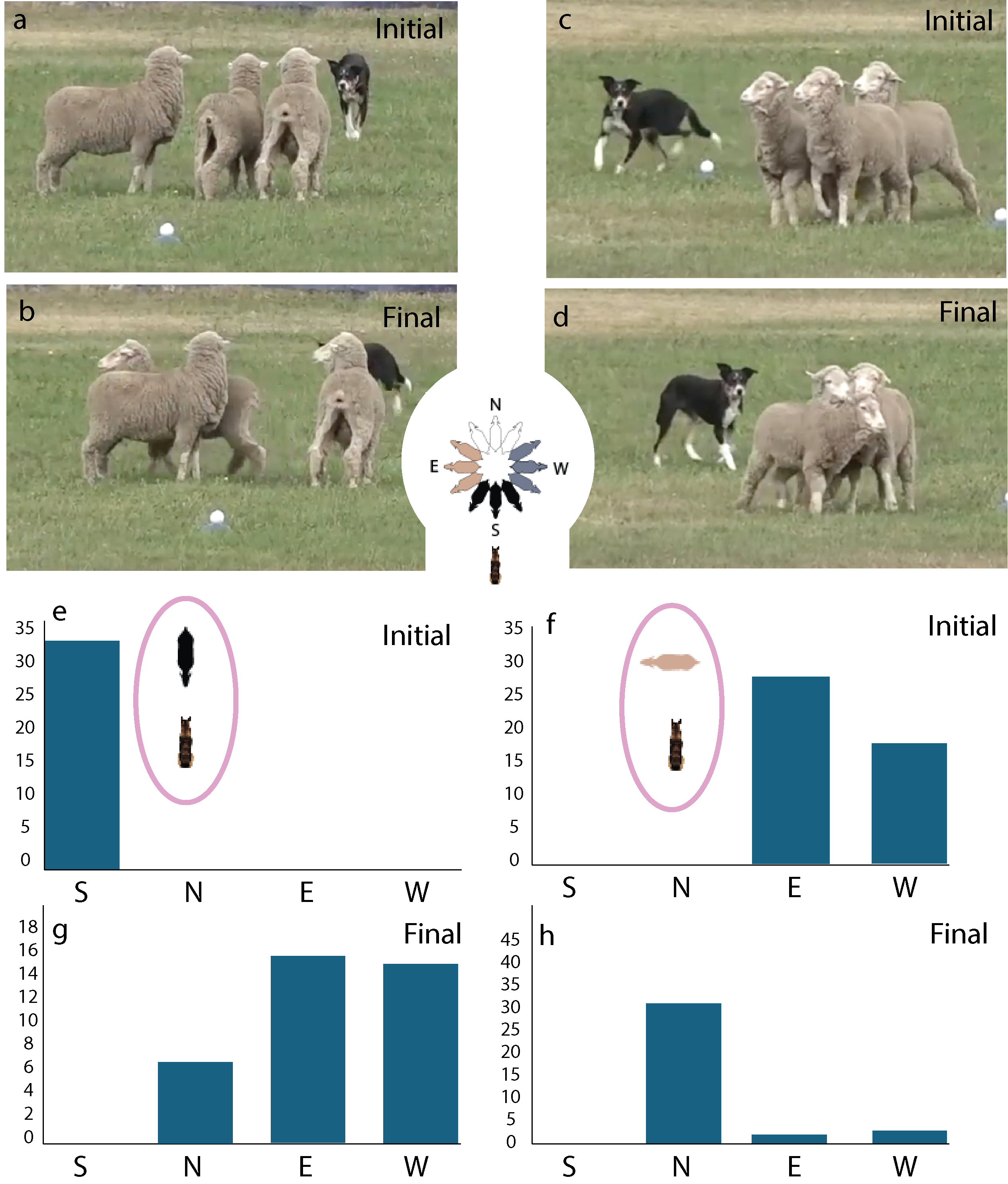}
    \caption{Orientation Rule: \textbf{a,b} Sheep orient in any random direction when the initial orientation is facing the dog (S). \textbf{c,d} When sheep are oriented perpendicular to the dog (E,W), they reorient to the direction away from the dog (N). We manually quantified this with 80 instances of stationary sheep flocks. \textbf{e,f} Histogram of initial states and \textbf{g,h} final states confirming our hypothesis. \textit{Videos Used in the analysis: 2024 National Sheep Dog Trial Championship series and 2023 National Sheep Dog Trial Championship series in YouTube channel Australian National Sheepdog Trial and A Way with the dogs 2 series on YouTube channel CSJ Specialist Canine Feeds. In this analysis, we found some sheep flocks show "milling"  where they start to rotate clockwise or anticlockwise at the same place without moving (For example, see https://youtu.be/E9MpU2hOMVY?si=gFeN3WAwqotixCsn\&t=146) . We exclude those videos from our analysis.}}
    \label{fig:SI3}
    \end{figure*}

\begin{table}[h!t]
\centering
\captionsetup{labelformat=simple, labelsep=period}
\caption*{Table S2: List of YouTube videos showing herding and shedding in light and heavy sheep} 
  {\fontsize{8}{10}\selectfont
\begin{tabular}{|m{5cm}|m{10cm}|}
    \hline
  
   \\ Herding Light Sheep & 
    \bulurl{https://youtu.be/r8BFg_Upw1w?si=bZHCir_nmBAjtEvT&t=331}\\
    & \bulurl{https://youtu.be/jbaYICYkQpU?si=DSOjetLygGrvSTGl&t=944}\\
    & 
    \\
    & \bulurl{https://youtu.be/jbaYICYkQpU?si=vY9klyMmpYsWQvE0&t=1609}\\
    & \\
    \hline
    Shedding Light Sheep & 
    \bulurl{https://youtu.be/r8BFg_Upw1w?si=BZFD4Ss5ZfJMbj0I&t=402}\\  
    & 
    \bulurl{https://youtu.be/cnPOXfUC8rc?si=EOQCUTpuLTdjD6pw&t=968}\\
    & 
    \\&
    \bulurl{https://youtu.be/jbaYICYkQpU?si=_gOGzZMFzUi6DyQ1&t=1820}\\
     & \\
    \hline
    Herding and Shedding Heavy Sheep & 
    \bulurl{https://www.youtube.com/live/u9lQ3NeHTVo?si=J6qhmejeIjRJkJst&t=1064}\\ 

    & \bulurl{https://www.youtube.com/live/u9lQ3NeHTVo?si=hQLi4PdhQ5ilRS6w&t=1963}\\
     & \\
    & \bulurl{https://www.youtube.com/live/8td_8L1qwIY?si=b6vn_RHfRxMs684J&t=353}\\
     & \\
    & Herding: \bulurl{https://youtu.be/jbaYICYkQpU?si=DSOjetLygGrvSTGl&t=944}\\
     & \\
    & Shedding: \bulurl{https://youtu.be/jbaYICYkQpU?si=mZKS_yZK_FeyuvaZ&t=1128}\\
     & \\
    \hline
    Herding Very Heavy Sheep & 
    \bulurl{https://www.youtube.com/watch?v=E9MpU2hOMVY&t=82s}\\ 
    & \bulurl{https://youtu.be/q4VRDGLJXa0?si=YRSsNtO9XXzz7hMO&t=248}\\ 
     & \\
    & \bulurl{https://youtu.be/q4VRDGLJXa0?si=FBlD-Cbwh2eHhOjc&t=1182}\\ 
     & \\
    & \bulurl{https://youtu.be/q4VRDGLJXa0?si=YMCIXpu0rqLFEXsK&t=2150}\\ 
     & \\
    & \bulurl{https://youtu.be/q4VRDGLJXa0?si=d_z0zBmJHABxlLQn&t=2456}\\
     & \\
    & \bulurl{https://youtu.be/E9MpU2hOMVY?si=6vlpFVhlrygwVjA7&t=69}\\ 
     & \\
    \hline 
   
\end{tabular}
 }
 \label{Table:S2}
\end{table}

\section{Derivation of the Transition Rates}

Let's consider a system of $N_s$ sheep where a sheep oriented in a direction $i \in (N,S,E,W)$ can transition to another direction $j\neq i$ with rates $\gamma$, $\alpha_{ik}$ and $\epsilon$ when influenced by other sheep, external stimulus positioned in direction $k$ or noise respectively. The rate of transition of the system from a state \(\Bar{x}\) to another state \(\Bar{y}\) can be written as

\begin{gather}
    Tr(\Bar{x} | \Bar{y}) = \sum_{i \in \{N, S, E, W\}} \sum_{j \neq i} T_{i \rightarrow j}
\end{gather}

$T_{i \rightarrow j}$, the rate of transition of an individual sheep from direction $i$ to $j$, is 

\begin{equation}
    T_{i \rightarrow j} = T_{i\rightarrow j}^{(\gamma)} + T_{i\rightarrow j}^{(\epsilon)} + \sum_k T_{i\rightarrow j}^{(\alpha_{ik})}
    \end{equation}

where $k$ is the position of the external stimuli. For our system, $k$ represents the positions of the dog for herding and the dog and the handler for shedding respectively, and   $T_{i\rightarrow j}^{(p)}$ is the transition rate from direction $i$ to direction $j$ with the influence $p \in ({\gamma, \epsilon, \alpha_{ik}})$. If a sheep in direction $i$ gets influenced by another sheep in direction $j$, it can reorient to direction $j$ the rate $\gamma$. The transition rate $T_{i\rightarrow j}^{(\gamma)}$ can be written as 

\begin{equation}
    T_{i\rightarrow j}^{(\gamma)} = \gamma x_i x_j
\end{equation}

where $x_i$ is the number of sheep oriented in direction $i$ and $x_j$ is the number of sheep oriented in direction $j$.

Similarly, if a sheep in direction $i$ is influenced by random noise, it can reorient to any of the remaining three direction with rate $\epsilon$. Hence the transition rate $T_{i\rightarrow j}^{(\epsilon)}$ becomes

\begin{equation}
    T_{i\rightarrow j}^{(\epsilon)} = \epsilon x_i
\end{equation}

The transition rate due to the stimulus $T_{i\rightarrow j}^{(\alpha_{ik})}$ depends on both the orientation of the sheep and the presence of the stimulus. If the stimulus approaches from the front, sheep panic and randomly reorient in a different direction. Therefore, a sheep oriented towards the stimulus will transition to any one of the three remaining directions with rate $\alpha_{kk}$, where $k$ is the direction of the stimulus.  However, if the stimulus approaches from the side, then the sheep reorient in the opposite direction from the stimulus. Hence, if the sheep is oriented perpendicular to the stimulus, it will transition to the opposite direction of the stimulus with rate $L\alpha_{kk}$. Where $L = \alpha_{ik}/\alpha_{kk}$ is the lightness of the sheep for $i\perp k$. When sheep is oriented in the opposite direction of the stimulus, it has no impact on the sheep. To summarize, the transition rate $T_{i\rightarrow j}^{(\alpha_{ik})}$ can be written as 

\begin{align}
    T_{i\rightarrow j}^{(\alpha_{ik})} &= 
    \begin{cases}
        \alpha_{kk} x_i & \text{if } i = k \in (N,S), \\
        L\alpha_{kk} x_i & \text{if } i \in (E,W) ;  k \in (N,S), \\
        0 & \text{otherwise}.
    \end{cases}
\end{align}


By incorporating Eqn 2,3 and 4 into Eq. 1, the transition rates between pair of directions $S$, $N$, $E$ and $W$ becomes

 \begin{align}
T_{S \rightarrow j} &= (\alpha_{SS} + \epsilon) x_S + \gamma x_S x_j, \\
T_{N \rightarrow j} &= (\alpha_{NN} + \epsilon) x_N + \gamma x_N x_j, \\
T_{E \rightarrow j} &= \left( L\left(\alpha_{SS} \delta_{j,N} + \alpha_{NN} \delta_{j,S} \right) +  \epsilon \right) x_E +  \gamma x_E x_j , \\
T_{W \rightarrow j} &= \left( L\left(\alpha_{SS} \delta_{j,N} + \alpha_{NN} \delta_{j,S} \right) +  \epsilon \right) x_W +  \gamma x_W x_j , \\
\end{align}

where $x_i$ is the number of sheep oriented in direction $i$ and $\delta_{m,n}$ is Kronecker delta function defined as 
\begin{equation*}
 \delta_{mn} =
\begin{cases}
    1 & \text{if } m = n, \\
    0 & \text{if } m \neq n.
\end{cases}
\end{equation*}

Since the master equation is not analytically tractable, we use Gillespie's algorithm to numerically simulate it as shown in the main text.

\section{Influence of noise $\epsilon$}

In the absence of other factors, the noise doesn't have a substantial influence on the ease of herding $E_h$ and ease of shedding $E_s$ since reach time ($\tau_{reach}$) and stay time ($\tau_{stay}$) in both herding and shedding decreases with an increase in $\epsilon$. However, in the presence of the influence of the dog and other sheep, the impact of noise becomes significant due to the presence of regimes where the dynamics only depends on $\epsilon$. Two such scenarios are staying time dynamics $\tau_{stay}$ in herding processes and reaching dynamics $\tau_{reach}$ in the shedding process of heavy sheep ($L = 0$). 

\begin{figure*}[htbp]
    \centering
    \includegraphics[width = .99\textwidth]{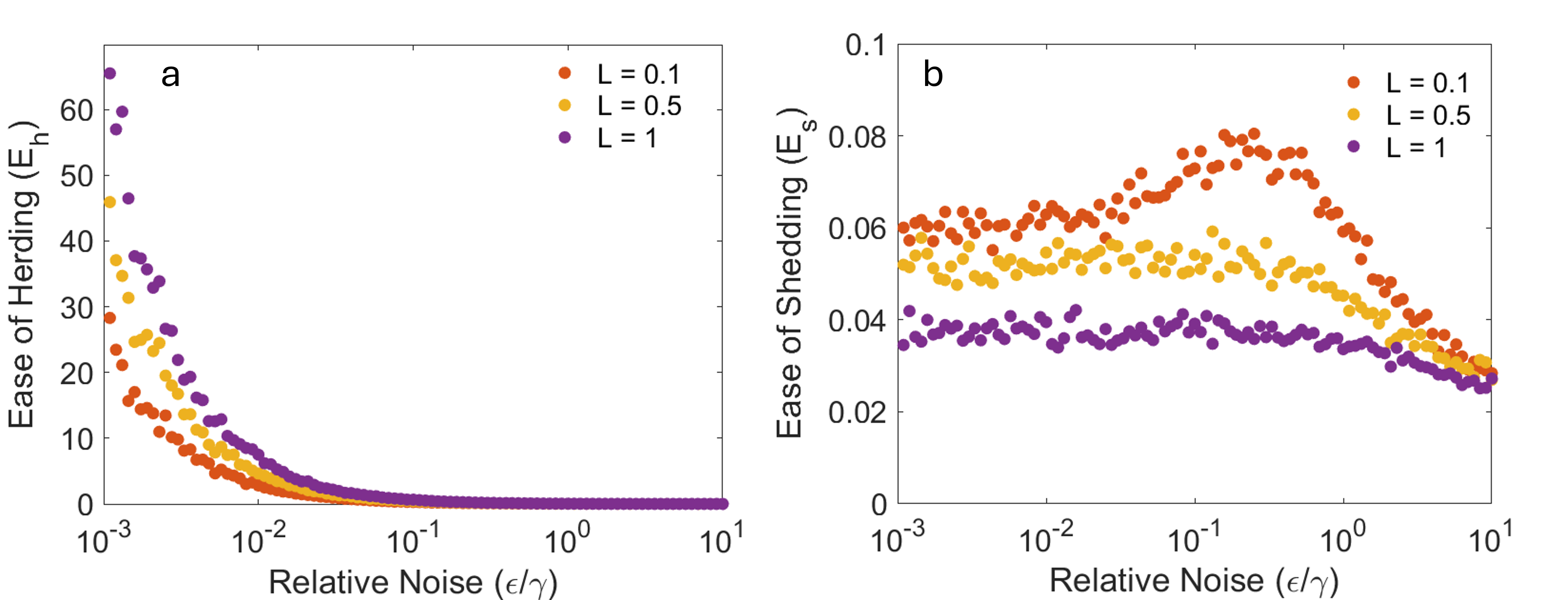}
    \caption{Effect of noise ($\epsilon$) on herding and shedding: a. In herding, for a given pressure and lightness, increasing $\epsilon$ decreases $E_h$. b. In shedding, for a given pressure, effect of $\epsilon$ of $E_s$ depends on lightness. For light sheep, $E_s$ monotonically decreases with an increase in $\epsilon$. However, for heavy sheep, $E_s$ is non-monotonic and is maximum for some intermediate $\epsilon$. Parameters: $P = 1$}
    \label{fig:SI4}
    \end{figure*}

In the herding process, when the sheep are in the herding state, neither the dog nor the other sheep have any influence on the individuals. This is apparent in Figure 3f of the main text where we observe no effect of lightness or pressure on the dynamics of $\tau_{stay}$. Therefore, for high $\alpha$ and $\gamma$, the noise has little impact on $\tau_{reach}$, but once the herding state is achieved, it can only be broken by the influence of noise with rate $\epsilon$. Hence, increasing $\epsilon$ is disadvantageous for herding Figure \ref{fig:SI4}a.

Similar to herding, increasing $\epsilon$ also decreases the stay time in shedding processes. However, for heavy sheep, if all the sheep are aligned in a perpendicular direction to the dog and the handler, the dog and the handler have minimal influence over the sheep. In such scenarios, the sheep can only switch orientation only due to the noise. Therefore, while the ease of shedding monotonically decreases with an increase in $\epsilon$ for light sheep, for heavy sheep, it is non-monotonic and is maximum for some intermediate $\epsilon$. Hence, the presence of noise is crucial for splitting groups of agents with non-isotropic responsiveness Figure \ref{fig:SI4}b.

\section{Unified phase diagram for sheep behavior}

Our analysis simplifies the sheepdog trials into a two-parameter system, i.e., pressure (P) and lightness (L) and we show the ease of herding and shedding can be estimated as functions of these two parameters. However, in reality, controlling a flock of sheep is much more complex because achieving the required pressure can be extremely challenging. This is because the flocking behavior in sheep has been reported to be a function of external stimuli (threat or allure). In the absence of any stimuli, sheep randomly wander around, and no interaction among the individuals is observed. On the other hand, for very strong stimuli such as an attack by a predator, sheep follow their individual instinct and tend to flee to save themselves from the threat. Only in the intermediate level of stimuli, collective dynamics is observed in sheep flocks which is caused by the interactions among the individuals in the group. Additionally, the intermediate regime may significantly vary depending on many external and internal parameters such as breed of sheep, age, flock size, etc.

To incorporate the above phenomenon in our model, we argue that  the transition rate due to the influence of other sheep $\gamma$ can be modeled as a negative quadratic function of the transition rate due to the external stimulus $\alpha$ as

\begin{equation}
    \gamma(\alpha) = a_0 + a_1 \alpha - a_2 \alpha^2
    \label{equation:7}
\end{equation}

where $a_0, a_1$ and $a_2$ are arbitrary constants. Note that our results in Figure 3 in the Main Text are still valid with this assumption. Additionally, pressure (P) is still a monotonic function of the stimulus (threat) $\alpha$, but not linear. Therefore, while in some regimes increasing the stimulus (threat) may not have any effect on the flock in some other regimes, a small increase in the stimulus (threat) may make the sheep panic and flee. Therefore, to achieve the optimal $P$, the dog-handler team must assess the relation between $\alpha$ and $\gamma$ for a given flock of sheep.

Introduction of Equation \ref{equation:7} in our model allows us to propose a complete phase diagram of the behavior space of sheep as a function of flock size ($N_s$) and specificity of the external stimulus which is defined as the ratio of the influence of the external stimuli and the influence of noise ($\alpha/\epsilon$). In this regard, we find three distinct regimes grazing, flocking, and fleeing. In grazing, the sheep are influenced by random stimuli that are not accounted for in our model such as food. Hence, in the absence of any external stimuli, the sheep distribute in the field and graze in random orientations. Therefore this regime is dominated by random noise. In flocking, the dynamics are dominated by the interactions between the sheep. Lastly, in fleeing, the sheep are mostly influenced by external stimuli, and instead of staying as a group, individuals decide to flee to save themselves from the threat. We define these regimes according to the dominating probability of sheep getting influenced by $\epsilon, \gamma$, or $\alpha$ respectively. We calculate the influence probabilities as

\begin{align}
Pr(\alpha) &= \frac{\alpha}{\alpha + N\gamma(\alpha) + \epsilon} \\
Pr(\gamma) &= \frac{N\gamma(\alpha)}{\alpha + N\gamma(\alpha) + \epsilon} \\
Pr(\epsilon) &= \frac{\epsilon}{\alpha + N\gamma(\alpha) + \epsilon}
\label{Equation:8}
\end{align}

where $\gamma(\alpha)$ is defined in Equation \ref{equation:7}. Figure 8 in the Main Text shows the three regimes as functions of group size ($N_s$) and specificity of the stimulus ($\alpha/\epsilon$) for an arbitrary parameter set. While changing parameters change the quantitative values, the qualitative behavior remains unaffected.   

\section{SI Videos and Table}

The following videos are provided as supplementary materials\\

    SI Video 1 . mp4: Example of small groups of sheep struggling to choose survival strategy, switching between solitary and collective
behaviors, creating unpredictability.\\

SI Video 2. mp4: Two steps of herding and shedding process: Herding and shedding involve two steps.: Initially, stationary sheep are gently nudged to induce directional change without causing panic. When the preferred direction is achieved, the pressure is intensified to prompt movement\\

SI Video 3. mp4: Four-direction representation of sheep orientation: We categorize orientations into 4 directions relative to the dog: directly facing, perpendicular left, perpendicular right, and facing away. The video demonstrates the four-direction representation.\\

SI Video 4. mp4: Herding and shedding of heavy and light sheep: While light sheep are easy to herd due to their high isotropy in responsiveness, they are very difficult to shed since they panic when sandwiched between the dog and the handler. On the other hand, heavy sheep are difficult to herd but easy to shed.\\ 

SI Video 5. mp4: Advantages of indecisiveness\\

SI Video 6. mp4: Video of light sheep herding\\

SI Video 7. mp4: Video of heavy sheep herding\\

SI Video 8. mp4: Video of light sheep shedding\\

SI Video 9. mp4: Video of heavy sheep shedding\\

SI Video 10. mp4: Demonstration of the rules of the model: Figure 2b and SI section 4c describe how a sheep's initial orientation with respect to dog's position changes the reorientation of the sheep. Video demonstrates an example of the result in figure 4. \\

SI Video 11. mp4: Comparison of ASA, ISA and LFSA: Demonstration of the dynamics for the trajectories in Figure 6b

\bibliographystyle{ieeetr}
\bibliography{Paperpile_Bib_Controlling_Noisy_Herds}